 \documentclass[final,1p,times]{elsarticle}

\usepackage{amssymb}
\usepackage{amsmath}
\usepackage{bm}

\usepackage{subcaption}
\usepackage{subfiles}

\usepackage{graphicx}
\usepackage{float}

\usepackage{tikz}
\usetikzlibrary{patterns}
\usepackage{geometry}

\usepackage{xcolor}

\usepackage{siunitx}

\usepackage[hidelinks,
            pdfauthor={Michael Mitterlindner, Maximilian Graber, Regina Kratzer, Markus Reichhartinger, Stefan Radl},
            pdftitle={Compartment Modelling of Multiphase Reactors using Unsupervised Clustering},
            pdfsubject={Chemical Engineering Journal},
            pdfkeywords={Computational Fluid Dynamics (CFD), Compartment
                            Modelling, Unsupervised Clustering, Reacting Multiphase Flow}]{hyperref}

\usepackage[automake,nomain,nogroupskip,abbreviations,symbols,stylemods={longbooktabs},sort=standard]{glossaries-extra}
\newglossary[slg]{greeksymbols}{gsy}{gbl}{Greek Characters}
\newglossary[slg]{latinsymbols}{lsy}{lbl}{Latin Characters}
\newglossary[slg]{operatorsymbols}{ops}{opl}{Operators}
\glsaddkey{unit}{\glsentrytext{\glslabel}}{\glsentryunit}{\GLsentryunit}{\glsunit}{\Glsunit}{\GLSunit}
\DeclareUnicodeCharacter{2212}{-}
\makeglossaries

\setabbreviationstyle{short}

\newabbreviation{2d}{2D}{Two-dimensional}

\newabbreviation{bc}{\text{BC}}{Boundary Condition}
\newabbreviation{cfd}{\text{CFD}}{Computational Fluid Dynamics}
\newabbreviation{clara}{\text{CLARA}}{Clustering Algorithm and Applications}
\newabbreviation{cm}{\text{CM}}{Compartment Model}
\newabbreviation{crn}{\text{CRN}}{Chemical Reactor Network}
\newabbreviation{cstr}{\text{CSTR}}{Continuously Stirred Tank Reactor}
\newabbreviation{cluster}{\text{c}}{Clusters}
\newabbreviation{comp}{\text{C}}{Compartment (\gls{cluster} times the number of phases)}
\newabbreviation{equil}{\text{eq}}{Equilibrium}
\newabbreviation{fluent}{\text{ANSYS Fluent}}{\href{https://www.ansys.com/products/fluids/ansys-fluent}{Ansys Fluent}}
\newabbreviation{gas}{\text{gas}}{Gaseous Phase}
\newabbreviation{global}{\text{glob}}{Global (over the whole setup)}
\newabbreviation{in}{\text{in}}{Input quantity}

\newabbreviation{interface}{\text{I}}{Interface (at the interface)}
\newabbreviation{local}{\text{local}}{Local (per compartment)}
\newabbreviation{liq}{\text{liq}}{Liquid Phase}
\newabbreviation{lbm}{\text{LBM}}{Lattice Bolzmann Method}
\newabbreviation{max}{\text{max}}{Maximal}
\newabbreviation{mass_trans}{\text{MT}}{Mass Transfer}
\newabbreviation{mix}{\text{MIX}}{Mixing}
\newabbreviation{mpc}{\text{MPC}}{Model Predictive Controll}
\newabbreviation{ode}{\text{ODE}}{Ordinary Differential Equation}
\newabbreviation{of}{OpenFOAM}
{Open Field Operation And Manipulation (\href{https://www.openfoam.com/news/main-news/openfoam-v2412}{v2412})}
\newabbreviation{out}{\text{out}}{Output quantity}

\newabbreviation{react}{\text{R}}{Reaction}
\newabbreviation{rmse}{\text{RMSE}}{Root Mean Squared Error}
\newabbreviation{rom}{\text{ROM}}{Reduced Order Model}
\newabbreviation{r2d2}{\text{R2D2-Phase}}{Reacting \gls{2d} two phase simulation case}

\newabbreviation{simvantage}{\text{SimVantage}}{\href{https://simvantage.com}{SimVantage}}
\newabbreviation{slsqp}{\text{SLSQP}}{\href{https://docs.scipy.org/doc/scipy/reference/optimize.minimize-slsqp.html}{Sequential Least SQuares Programming}}
\newabbreviation{source}{\text{s}}{Source therm}
\newabbreviation{superficial}{\text{S}}{Superficial}
\newabbreviation{tot}{\text{tot}}{Total}
\newabbreviation{phi_phase}{\ensuremath{\varphi}}{Phase}
\glsnoexpandfields

\glsxtrnewsymbol[type=latinsymbols,
    description={Example variable can be a scalar or vector quantity},
	unit=\unit{[-]}]
    {a}
    {\ensuremath{a}}
    
\glsxtrnewsymbol[type=latinsymbols,
    description={Blending factor for the Superposition of a linear and a quadratic flow profile},
	unit=\unit{[-]}]
    {lqfp_blend}
    {\ensuremath{a_{blend}}}

\glsxtrnewsymbol[type=latinsymbols,
    description={Intermediate parameter for solveing the two-phase masstransfer equation},
	unit=\unit{[\kilo\mole\per\meter\tothe{4}]}]
    {inter_para_tpmte}
    {\ensuremath{b}}

\glsxtrnewsymbol[type=latinsymbols,
    description={Dimensionless intermediate parameter for solveing the two-phase masstransfer equation},
	unit=\unit{[-]}]
    {inter_para_tpmte_dimless}
    {\ensuremath{b^*}}

\glsxtrnewsymbol[type=latinsymbols,
    description={Concentration},
	unit=\unit{[\kilo\mol\per\cubic\meter]}]
    {conc}
    {\ensuremath{c}}
    
\glsxtrnewsymbol[type=latinsymbols,
    description={Concentration vector of reactors $\mathbb{R}^{\gls{number}_{\gls{comp}} \times 1}$},
	unit=\unit{[\kilo\mol\per\cubic\meter]}]
    {conc_vec}
    {\ensuremath{\bm{c}}}

\glsxtrnewsymbol[type=latinsymbols,
    description={Cell labels},
	unit=\unit{[-]}]
    {Cell_label}
    {\ensuremath{C}}
    
\glsxtrnewsymbol[type=latinsymbols,
    description={Diameter},
	unit=\unit{[\meter]}]
    {diameter}
    {\ensuremath{d}}
    
\glsxtrnewsymbol[type=latinsymbols,
    description={Diffusion coefficient},
	unit=\unit{[\square\meter\per\second]}]
    {diff_coeff}
    {\ensuremath{D}}

\glsxtrnewsymbol[type=latinsymbols,
    description={Damköhler number (reaction time scales / residence time)},
	unit=\unit{[-]}]
    {Da}
    {\ensuremath{\text{Da}}}

\glsxtrnewsymbol[type=latinsymbols,
    description={Height},
	unit=\unit{[-]}]
    {height}
    {\ensuremath{h}}

\glsxtrnewsymbol[type=latinsymbols,
    description={Hatta number (rate of reaction / diffusion)},
	unit=\unit{[-]}]
    {hatta_number}
    {\ensuremath{\text{Ha}}}

\glsxtrnewsymbol[type=latinsymbols,
    description={Index (used for species)},
	unit=\unit{[-]}]
    {index}
    {\ensuremath{i}}

\glsxtrnewsymbol[type=latinsymbols,
    description={Turbulent Kinetic Energy},
	unit=\unit{[\square\meter\per\square\second]}]
    {tke}
    {\ensuremath{k^{turb}}}

\glsxtrnewsymbol[type=latinsymbols,
    description={Reaction rate constant 1st order},
	unit=\unit{[1/\second]}]
    {rrc1O}
    {\ensuremath{k^{1^{st}}}}

\glsxtrnewsymbol[type=latinsymbols,
    description={Reaction rate constant},
	unit=\unit{[variable]}]
    {rrc}
    {\ensuremath{k^{react}}}

\glsxtrnewsymbol[type=latinsymbols,
    description={Reaction rate constant 2nd order},
	unit=\unit{[\cubic\meter\per(\kilo\mole \ \second)]}]
    {rrc2O}
    {\ensuremath{k^{2^{nd}}}}

\glsxtrnewsymbol[type=latinsymbols,
    description={Reaction rate constant 0.5th order},
	unit=\unit{[(\kilo\mole\per\cubic\meter)^{0.5}\second^{-1}]}]
    {rrc05O}
    {\ensuremath{k^{0.5^{th}}}}

\glsxtrnewsymbol[type=latinsymbols,
    description={Volumetric mass transfer coefficient (total)},
	unit=\unit{[1\per\second]}]
    {kLa}
    {\ensuremath{k_La}}

\glsxtrnewsymbol[type=latinsymbols,
    description={Volumetric mass transfer coefficient (total) per compartment $\mathbb{R}^{\gls{number}_{\gls{comp}} \times 1}$},
	unit=\unit{[1\per\second]}]
    {kLa_vec}
    {\ensuremath{\bm{k_La}}}
    
\glsxtrnewsymbol[type=latinsymbols,
    description={Volumetric mass transfer coefficient (per liquid volume)},
	unit=\unit{[1\per\second]}]
    {kLa_liq}
    {\ensuremath{k_La_{\gls{liq}}}}

\glsxtrnewsymbol[type=latinsymbols,
    description={Channel length (single phase)},
	unit=\unit{[\meter]}]
    {length_channel}
    {\ensuremath{L}}

\glsxtrnewsymbol[type=latinsymbols,
    description={Reference length (two phase)},
	unit=\unit{[\meter]}]
    {length_ref}
    {\ensuremath{L_{ref}}}

\glsxtrnewsymbol[type=latinsymbols,
    description={Lewis number},
	unit=\unit{[-]}]
    {lewis_number}
    {\ensuremath{\text{Le}}}

\glsxtrnewsymbol[type=latinsymbols,
    description={Reaction order},
	unit=\unit{[-]}]
    {react_order}
    {\ensuremath{m}}

\glsxtrnewsymbol[type=latinsymbols,
    description={Number},
	unit=\unit{[-]}]
    {number}
    {\ensuremath{n}}

\glsxtrnewsymbol[type=latinsymbols,
    description={Molar  rate per unit total volume},
	unit=\unit{[\kilo\mole\per(\cubic\meter \ \second)]}]
    {mol_trans_rate_per_vol}
    {\ensuremath{R_{vol}}}
    
\glsxtrnewsymbol[type=latinsymbols,
    description={Molar rate},
	unit=\unit{[\kilo\mole\per\second]}]
    {mol_rate}
    {\ensuremath{R}}

\glsxtrnewsymbol[type=latinsymbols,
    description={Molar weight},
	unit=\unit{[\kilo\gram\per\kilo\mole]}]
    {mol_weight}
    {\ensuremath{MW}}

\glsxtrnewsymbol[type=latinsymbols,
    description={Pressure},
	unit=\unit{[\pascal]}]
    {pressure}
    {\ensuremath{p}}

\glsxtrnewsymbol[type=latinsymbols,
    description={Prantl number},
	unit=\unit{[-]}]
    {prantl_number}
    {\ensuremath{\text{Pr}}}

\glsxtrnewsymbol[type=latinsymbols,
    description={Rate of reaction},
    unit={\unit{[\kilo\mole\per(\cubic\meter\second)]}}]
    {rate_react}
    {\ensuremath{r}}

\glsxtrnewsymbol[type=latinsymbols,
    description={Turbulent Schmidt number},
    unit={\unit{[-]}}]
    {schmidt_number_turbulent}
    {\ensuremath{\text{Sc}_t}}
    
\glsxtrnewsymbol[type=latinsymbols,
    description={Time},
	unit=\unit{[\second]}]
    {time}
    {\ensuremath{t}}

\glsxtrnewsymbol[type=latinsymbols,
    description={Temperature},
	unit=\unit{[\kelvin]}]
    {temp}
    {\ensuremath{T}}

\glsxtrnewsymbol[type=latinsymbols,
    description={Velocity},
	unit=\unit{[\meter\per\second]}]
    {velo}{\ensuremath{u}}

\glsxtrnewsymbol[type=latinsymbols,
    description={Velocity (vector)},
	unit=\unit{[\meter\per\second]}]
    {velo_vec}
    {\ensuremath{\bm{u}}}

\glsxtrnewsymbol[type=latinsymbols,
    description={Volume},
	unit=\unit{[\cubic\meter]}]
    {Vol}
    {\ensuremath{V}}

\glsxtrnewsymbol[type=latinsymbols,
    description={Vector of compartment volumes $\mathbb{R}^{\gls{number}_{\gls{comp}} \times 1}$},
	unit=\unit{[\cubic\meter]}]
    {Vol_vec}
    {\ensuremath{\bm{V}}}

\glsxtrnewsymbol[type=latinsymbols,
    description={Weight fraction},
	unit=\unit{[-]}]
    {weight_frac}
    {\ensuremath{w}}

\glsxtrnewsymbol[type=latinsymbols,
    description={Molar flow rate},
	unit=\unit{[\kilo\mole\per\second]}]
    {mol_flow}
    {\ensuremath{X}}

\glsxtrnewsymbol[type=latinsymbols,
    description={Molar flux},
	unit=\unit{[\kilo\mole\per(\square\meter\second)]}]
    {mol_flux}
    {\ensuremath{\dot X}}

\glsxtrnewsymbol[type=latinsymbols,
    description={Gravitation 9.81},
	unit=\unit{[\meter\per\square\second]}]
    {gravity}
    {\ensuremath{\text{g}}}

\glsxtrnewsymbol[type=latinsymbols,
    description={Henry's constant (our case dimension less)},
	unit=\unit{[-]}]
    {henry_const}
    {\ensuremath{\text{H}}}

\glsxtrnewsymbol[type=latinsymbols,
    description={Universal gas constant 8314.46  },
	unit=\unit{[\joule\per(\kilo\mole\kelvin)]}]
    {gas_const}
    {\ensuremath{\text{R}}}

\glsxtrnewsymbol[type=latinsymbols,
    description={Relative Tolerance},
	unit=\unit{[-]}]
    {rel_tol}
    {\ensuremath{tol}}
\glsnoexpandfields

\glsxtrnewsymbol[type=greeksymbols,
    description={Phase/volume fraction},
	unit=\unit{[-]}]
    {vol_frac}
    {\ensuremath{\alpha}}
    
\glsxtrnewsymbol[type=greeksymbols,
    description={Phase/volume fraction $\mathbb{R}^{\gls{number}_{\gls{cluster}} \times 1}$},
	unit=\unit{[-]}]
    {vol_frac_vec}
    {\ensuremath{\bm{\alpha}}}

\glsxtrnewsymbol[type=greeksymbols,
    description={Error},
	unit=\unit{[-]}]
    {error}
    {\ensuremath{\epsilon}}

\glsxtrnewsymbol[type=greeksymbols,
    description={Hold-up},
	unit=\unit{[-]}]
    {hold_up}
    {\ensuremath{\varepsilon}}

\glsxtrnewsymbol[type=greeksymbols,
    description={Spacial coordinate},
	unit=\unit{[m]}]
    {coord}
    {\ensuremath{\xi}}

\glsxtrnewsymbol[type=greeksymbols,
    description={Spacial coordinate (vector)},
	unit=\unit{[m]}]
    {coord_vec}
    {\ensuremath{\bm{\xi}}}

\glsxtrnewsymbol[type=greeksymbols,
    description={Mole fraction},
	unit=\unit{[-]}]
    {mol_frac}
    {\ensuremath{\chi}}

\glsxtrnewsymbol[type=greeksymbols,
    description={Regularization therm},
	unit=\unit{[-]}]
    {regularization_therm}
    {\ensuremath{\lambda}}
    
\glsxtrnewsymbol[type=greeksymbols,
    description={Dynamic viscosity},
	unit=\unit{[\pascal \ \second]}]
    {dynamic_visc}
    {\ensuremath{\mu}}
    
\glsxtrnewsymbol[type=greeksymbols,
    description={Stoichiometric coefficients},
	unit=\unit{[-]}]
    {stoich_coeff}
    {\ensuremath{\nu}}
    
\glsxtrnewsymbol[type=greeksymbols,
    description={Density},
	unit=\unit{[\kilo\gram\per\cubic\meter]}]
    {density}
    {\ensuremath{\rho}}

\glsxtrnewsymbol[type=greeksymbols,
    description={Surface tension},
	unit=\unit{[\newton\per\meter]}]
    {surface_tension}
    {\ensuremath{\sigma}}
    
\glsxtrnewsymbol[type=greeksymbols,
    description={Volumetric flow rate (scalar)},
	unit=\unit{[\cubic\meter\per\second]}]
    {vol_flow}
    {\ensuremath{\phi}}

\glsxtrnewsymbol[type=greeksymbols,
    description={Volumetric flow rate (vector) (in/outflows) $\mathbb{R}^{\gls{number}_{\gls{cluster}} \times 1}$},
	unit=\unit{[\cubic\meter\per\second]}]
    {vol_flow_vec}
    {\ensuremath{\bm{\phi}}}

\glsxtrnewsymbol[type=greeksymbols,
    description={Volumetric flow rate matrix (flow rates between compartments $\mathbb{R}^{\gls{number}_{\gls{cluster}} \times \gls{number}_{\gls{cluster}}}$)},
	unit=\unit{[\cubic\meter\per\second]}]
    {vol_flow_mat}
    {\ensuremath{\bm{\Phi}}}

\glsxtrnewsymbol[type=greeksymbols,
    description={Time scale},
	unit=\unit{[\second]}]
    {time_scale}
    {\ensuremath{\tau}}

\glsnoexpandfields

\newglossaryentry{dimlessquant}{
  type=operatorsymbols,
  name={$\gls{a} ^ {*}$},
  description={Dimensionless quantity}
}

\newglossaryentry{spatialavg}{
  type=operatorsymbols,
  name={$\langle \gls{a} \rangle$},
  description={Spatial average}
}

\newglossaryentry{temporalavg}{
  type=operatorsymbols,
  name={$\overline{\gls{a}}$},
  description={Temporal average}
}

\newglossaryentry{outer}{
  type=operatorsymbols,
  name={$\otimes$},
  description={Outer product}
}

\newglossaryentry{hadamard_product}{
  type=operatorsymbols,
  name={$\circ$},
  description={Hadamard (element-wise) product}
}

\newglossaryentry{hadamard_division}{
  type=operatorsymbols,
  name={$\oslash$},
  description={Hadamard (element-wise) division}
}

\newglossaryentry{diagvec}{
  type=operatorsymbols,
  name={$\bm{v} = \mathrm{diag}(\bm{M})$},
  description={Vector $\bm{v}$ containing the main diagonal of matrix $\bm{M}$}
}

\newglossaryentry{diagmat}{
  type=operatorsymbols,
  name={$\bm{M} = \mathrm{diag}(\bm{v})$},
  description={Matrix $\bm{M}$ with the elements of vector $\bm{v}$ on the main diagonal}
}

\journal{Chemical Engineering Journal}

\begin{document}

\begin{frontmatter}



\title{Compartment Modelling of Multiphase Reactors using Unsupervised Clustering} 

\author[IPPT]{Michael Mitterlindner}
\author[IRT]{Maximilian Graber}
\author[BIOTE]{Regina Kratzer}
\author[IRT]{Markus Reichhartinger}
\author[IPPT]{Stefan Radl\corref{cor1}}

\cortext[cor1]{Corresponding author.}
\ead{radl@tugraz.at}

\affiliation[IPPT]{organization={Institute of Process and Particle Engineering (IPPT), Graz University of Technology},
            addressline={Inffeldgasse 13/III}, 
            city={Graz},
            postcode={8010}, 
            state={Styria},
            country={Austria}}

\affiliation[IRT]{organization={Institute of Automation and Control (IRT), Graz University of Technology},
            addressline={Inffeldgasse 21/B/I}, 
            city={Graz},
            postcode={8010}, 
            state={Styria},
            country={Austria}}

\affiliation[BIOTE]{organization={Institute of Biotechnology and Biochemical Engineering (BIOTE), Graz University of Technology},
            addressline={Petersgasse 10-12/I}, 
            city={Graz},
            postcode={8010}, 
            state={Styria},
            country={Austria}}

\begin{abstract}

   Detailed \glsentrylong{cfd} (\gls{cfd}) simulations are too computationally expensive for the real-time control and design optimization of multiphase flow reactors. To address these limitations, we introduce \gls{clara}, a software toolbox that automates the generation of \glsentrylong{cm}s (\gls{cm}) via the unsupervised clustering of \gls{cfd} data. Unlike previous studies, our toolbox enables the modelling of multiphase phenomena and interphase mass transfer within each compartment. \gls{clara} employs unsupervised clustering algorithms, graph reassignment, and optimization routines to ensure mass conservation and spatial connectivity across all compartments. Verification studies utilizing analytical benchmarks and reactive multiphase \gls{cfd} simulations demonstrate that the \glspl{cm} produced by \gls{clara} accurately reproduce reactor performance and spatial species distributions. The significantly reduced computational demand of \glspl{cm} compared to full \gls{cfd} models enables the optimal control of multiphase reactors and facilitates their rational design and optimization.

\end{abstract}


\end{frontmatter}



\newpage


\section{Introduction} \label{sec:introduction}
\subsection{Background}

The operation, design and optimization of industrial-scale reactors, ranging from chemical synthesis and polymerization to mineral carbonation and biotechnology, rely heavily on managing spatial heterogeneities. In large-scale vessels, non-ideal mixing inevitably leads to gradients in critical process parameters such as substrate concentration, temperature, pH, and dissolved gas levels \cite{laborda_unsupervised_2025, jamshidzadeh_cfd-guided_2025}. These inhomogeneities can severely impact reaction rates, product yield, microbial physiology, and overall process stability \cite{nadal-rey_development_2021, schwarz_cfd-based_2023}. The severity of these effects is strongly coupled to the reaction kinetics: while first- and second-order reactions are often relatively insensitive to local concentration variations, fractional-order kinetics, as encountered, for example, in oxygen-limited bioprocesses, exhibit a pronounced dependence on local concentration gradients \cite{danckwerts_absorption_1950}, making spatial resolution of the reactor indispensable.

While \glsentrylong{cfd} (\gls{cfd}) provides high-resolution insights into flow fields and transport phenomena, it remains computationally prohibitive when simulating long process durations, such as fed-batch fermentations which may last days or weeks \cite{le_nepvou_de_carfort_automatic_2024, maldonado_de_leon_dynamic_2025}. Furthermore, the computational cost of full \gls{cfd} makes it impractical for real-time applications like digital twins or Model Predictive Control (MPC) \cite{le_nepvou_de_carfort_modeling_2026, lambauer_automatic_2023}. \glsentrylong{cm}s (\gls{cm}), also known as \glsentrylong{crn}s (\gls{crn}) or Network-of-Zones models, which are a special category of \glsentrylong{rom}s (\gls{rom}), offer a necessary compromise. By simplifying hydrodynamics and dividing the reactor into interconnected, ideally mixed zones, \gls{cm}s reduce computational load by orders of magnitude, potentially accelerating simulations by factors of over 2000 compared to full \gls{cfd}, while retaining critical information regarding spatial heterogeneity \cite{le_nepvou_de_carfort_automatic_2024, savarese_machine_2023}. The low computational cost of this approach makes it an excellent candidate for surrogate modelling in \gls{mpc} and digital twin applications. These applications inherently require rapid, repeated evaluations over long time horizons. Furthermore, by providing higher spatial resolution than conventional zero-dimensional models, compartment models can be effectively integrated into multi-agent systems \cite{rupprecht_multi-agent_2026} or used in simulation ontologies \cite{zhang_semantic_2026}.

\subsection{State of the Art and Gaps in Knowledge}

The development of compartment models has evolved through distinct generations. Early approaches relied on manual zoning based on expert intuition or limited experimental data \cite{le_nepvou_de_carfort_flow-informed_2026}. Second-generation models utilized \gls{cfd} velocity fields to inform fluxes between manually defined zones or structured grids \cite{delafosse_cfd-based_2014}. Currently, the field is moving toward automated third- and fourth-generation compartment models that employ unsupervised machine learning to partition the reactor geometry. Recent advancements include the use of k-means and hierarchical agglomerative clustering to group \gls{cfd} cells based on flow variables, residence time, or chemical concentration \cite{laborda_unsupervised_2025, le_nepvou_de_carfort_flow-informed_2026}, combined with graph reassignment algorithms that enforce the spatial connectivity of the resulting clusters \cite{savarese_machine_2023}. These techniques have been applied across various domains, including bioreactor \cite{promma_coupled_2024}, combustion \cite{savarese_machine_2023}, polymerization \cite{schwarz_cfd-based_2023}, and mineral carbonation applications \cite{kim_modeling_2020}. Recently, such automated clustering approaches have also been identified as crucial for simplifying the complex hydrodynamics of large-scale gas fermenters into computationally efficient networks of ideally mixed compartments \cite{puiman_dos_2025}.

Despite these advancements, significant knowledge gaps remain. Automated clustering methods, while effective for single-phase flows, are rarely extended to multiphase applications. Studies on gas-liquid-solid reactors and multiphase loop reactors still rely on manual spatial division or fixed grid-based approaches to handle phase interactions \cite{weber_cfd_2019, kim_modeling_2020, promma_coupled_2024}. Furthermore, existing methodologies are often tightly coupled to a specific commercial software, such as \gls{fluent} \cite{savarese_machine_2023}, or depend on proprietary in-house software packages \cite{le_nepvou_de_carfort_flow-informed_2026}. No open-source toolbox currently exists that can process data from multiple \gls{cfd} solvers, e.g., from \gls{of} and \gls{fluent}, to automatically generate compartment models for multiphase reactor flows.

Mass conservation presents a further challenge. Interpolation errors introduced during compartmentalization can lead to unclosed mass balances, compromising numerical stability over extended simulations. While recent work addressed this by mapping \gls{cfd} data onto automatically generated grids and applying iterative correction schemes \cite{le_nepvou_de_carfort_automatic_2024}, mass conservation remains particularly non-trivial for Lattice Boltzmann Method (LBM) simulations, which are increasingly adopted in bioreactor modeling. Beyond mass conservation, the choice of clustering algorithm critically shapes the topology of the resulting compartment network. Spherical methods such as k-means produce compact, isotropic zones, whereas hierarchical agglomerative clustering yields elongated, anisotropic structures \cite{murtagh_algorithms_2012}. It remains unclear which approach is better suited to different mixing regimes, or whether this choice has a measurable effect on \gls{cm} accuracy for reactive multiphase systems.

A related limitation concerns the breakdown of the well-mixed assumption at high spatial resolution. In turbulent flows, sub-grid dispersion across compartment boundaries is not captured by net convective fluxes alone \cite{delafosse_cfd-based_2014}, which can introduce artificial concentration gradients and narrow the useful range of compartment numbers. The impact of this effect on reactive multiphase systems has not been systematically quantified. Finally, validation efforts typically benchmark reduced-order predictions against full \gls{cfd} results or experimental tracer curves \cite{tajsoleiman_cfd_2019}, without first verifying the mathematical correctness of the underlying compartmentalization algorithm against known analytical solutions. Such verification is essential before applying any automated framework to complex turbulent flow scenarios.

\subsection{Goals}

To address these limitations, we introduce \gls{clara}, a software toolbox that automates the generation of \glsentrylong{cm} (\gls{cm}) via unsupervised clustering of \gls{cfd} data. Unlike previous methods, \gls{clara} is designed to explicitly model multiphase interactions and interface mass transfer within an automatically generated, arbitrarily shaped compartment network. While our mass transfer model is validated for gas dissolving in a liquid, the software architecture is designed to be applicable to a wide range of multiphase applications.

The present work pursues four overarching objectives: (i) Methodological Development: to implement a robust, mass-conserving compartmentalization framework. (ii) Mathematical Verification: to rigorously verify the \gls{clara} implementation against analytical solutions for both single-phase reactive channel flows and two-phase flows with mass transfer, ensuring the mathematical correctness of the procedure. (iii) Application and Sensitivity Analysis: to apply the framework to reactive two-phase \gls{of} simulations of a gas-liquid bubble column with non-linear reaction kinetics, quantifying the impact of turbulent sub-grid mixing and determining the optimal cluster count (and hence the complexity of the CM) to predict species distribution. (iv) Practical Guidelines and Accessibility: to derive clear heuristics for clustering algorithm selection under varying mixing conditions, and to provide the resulting open-source toolbox with broad compatibility across various \gls{cfd} solvers (including \gls{of}, \gls{fluent}, and the \gls{lbm} solver developed by \gls{simvantage}). \\

The remainder of our manuscript is structured as follows. Section \ref{sec:theoretical_models} establishes the theoretical framework, detailing the compartment modelling approach for convective transport and interphase mass transfer, alongside the configuration for the reactive two-phase \gls{cfd} reference simulation. Section \ref{sec:numerical_methods} presents the numerical methods, outlining the used clustering algorithms, as well as the flow optimization procedure designed to enforce mass conservation across the network. Section \ref{sec:results} discusses the results, beginning with a mathematical verification of the framework against analytical benchmarks, followed by an in depth analysis of the quarter-gassed bubble column scenarios to evaluate the impact of feature selection, non-linear reaction kinetics, and turbulent mixing on model accuracy. Finally, Section \ref{sec:conclustion_outlook} summarizes the key conclusions, offers practical guidelines for algorithm selection, and outlines directions for future research

\section{Theoretical Models} \label{sec:theoretical_models}


\subsection{Compartment Model} \label{sec:compartment_modeling}
\gls{cm} are a class of \glsentrylong{rom}s (\gls{rom}) that decompose the reactor domain into a finite number of ideally mixed zones, each described by a single averaged concentration. They are closely related to \glsentrylong{crn}s (\gls{crn}), though the \gls{cm} framework is broader: it is not restricted to reactive systems but naturally accommodates interphase mass transfer, as demonstrated here for gas-liquid systems. The derivation below is presented for gas-liquid transfer, but the formulation is equally applicable to other mass transfer scenarios, e.g.\ solid-gas or liquid-liquid transfer in extraction applications.

The concentration balance for each compartment in a \gls{cm} of $\gls{number}$ \gls{cstr}s is given by \autoref{eq:dcdt_compartmentModel}:

\begin{equation}
\frac{d\gls{conc_vec}}{d\gls{time}} =
\underbrace{\mathbf{s}_{\text{conv}}(\gls{conc_vec}, \gls{vol_flow_mat}, \gls{vol_flow_vec}^{\gls{bc}})}_{\text{convective transport}}
\; + \;
\underbrace{\mathbf{s}_{\text{react}}(\gls{conc_vec},  \gls{stoich_coeff}, \gls{rrc})}_{\text{reaction}}
\; + \;
\underbrace{\mathbf{s}_{\text{\gls{mass_trans}}}(\gls{conc_vec}, \gls{kLa}, \gls{henry_const}^*)}_{\text{mass transfer}}
\label{eq:dcdt_compartmentModel}
\end{equation}

These concentrations represent the phase-averaged volumetric concentration in each compartment, its computation is crucial and, therefore, it is described in more detail below.
It is important to note that concentrations are defined per phase volume, not with respect to the total reactor volume. Clustering is performed such that each phase occupies the same total spatial regions, i.e.\ the cluster boundaries are identical for all phases, but the phase volume differs in each region. Consequently, a two-phase system with $N$ clusters yields $2N$ compartments in total: $N$ for the liquid phase and $N$ for the gas phase.

\subsubsection{Convective Transport}

The \autoref{eq:matrixFlowCalculation} shows how the convective Transport is modelled. The inflow term ($\gls{vol_flow_mat}^\top \gls{conc_vec}_{\gls{index}}$) calculates the flow entering a compartment by multiplying the incoming volumetric flows by the concentrations of the compartments. Conversely, the outflow term calculates the flow leaving; because each compartment is ideally mixed, all exiting fluid uniformly carries the compartment's own internal concentration. To compute this efficiently using linear algebra, the operation $\gls{vol_flow_mat}\bm{1}$ sums the rows of the flow matrix to yield a single vector representing the total  flow leaving each compartment. The $\text{diag}$ operator then places these total outgoing flows along the diagonal of a new square matrix. Finally, multiplying this diagonal matrix by the concentration vector ($\gls{conc_vec}_{\gls{index}}$) accurately pairs and multiplies each compartment's total outgoing flow by its own internal concentration. We note in passing that "$\circ$" denotes the element-wise multiplication.
\begin{equation}
\mathbf{s}_{\text{conv}} =
\frac{1}{\gls{Vol_vec}} \circ
\left(
\underbrace{\gls{vol_flow_mat}^\top \gls{conc_vec}_{\gls{index}}}_{\text{inflow}}
\; - \;
\underbrace{\text{diag} \left(\gls{vol_flow_mat}\bm{1}\right) \gls{conc_vec}_{\gls{index}}}_{\text{outflow}}
\; + \;
\underbrace{\gls{bc}}_{\text{boundary terms}}
\right)
\label{eq:matrixFlowCalculation}
\end{equation}

In the following, each term is discussed in greater detail.

\paragraph{Internal Flows}

The flow matrix $\gls{vol_flow_mat}$ encodes the volumetric flow rates between compartments and is assembled directly from the \gls{cfd} cell-to-cell flow field (i.e., the volumetric flow between two cells). In \gls{of}, each internal face is associated with an owner cell and a neighbour cell. We use a signed volumetric flow rate $\gls{vol_flow}$ whose sign convention follows the outward face normal of the owner cell (i.e., positive flow rates leave the owner, negative flow rates enters it). For every internal face whose owner and neighbour belong to different clusters $j$ and $k$, the corresponding entry of $\gls{vol_flow_mat}$ is incremented, respecting the sign of $\gls{vol_flow}$. Faces shared by cells of the same cluster contribute only to the internal recirculation within that compartment and are therefore discarded. The construction rule is given in \autoref{eq:flowMatDefinition}, with an example of the construction of a three compartment flow matrix:

\begin{equation}
\gls{vol_flow_mat}=
\begin{cases}

\gls{vol_flow}_{j \rightarrow k} ,   & \text{if } \gls{vol_flow} > 0 \text{ and } \gls{Cell_label}_{owner} \neq \gls{Cell_label}_{neighbor} \\
-\gls{vol_flow}_{j \rightarrow k} ,  & \text{if } \gls{vol_flow} < 0  \text{ and } \gls{Cell_label}_{owner} \neq \gls{Cell_label}_{neighbor}  \\
0,                                     & \text{else}  \\
\end{cases}
\rightarrow
\begin{bmatrix}
0 & \gls{vol_flow}_{0 \rightarrow 1} & \gls{vol_flow}_{0 \rightarrow 2} \\
\gls{vol_flow}_{1 \rightarrow 0} & 0 & \gls{vol_flow}_{1 \rightarrow 2} \\
\gls{vol_flow}_{2 \rightarrow 0} & \gls{vol_flow}_{2 \rightarrow 1} & 0
\end{bmatrix}
\label{eq:flowMatDefinition}
\end{equation}

The diagonal is zero by construction (no self-flow). Each entry $\gls{vol_flow_mat}_{jk}$ represents the volumetric flow rate from compartment $j$ to compartment $k$, all entries are positive by construction. The matrix on the right-hand side of \autoref{eq:flowMatDefinition} illustrates the structure of $\gls{vol_flow_mat}$ for a three-compartment system.

\paragraph{Boundary Terms}
When defining boundary conditions for system modelling, concentrations are typically established as either fixed values or dynamic states derived directly from the compartment itself. Fixed boundary concentrations ($\gls{conc_vec}^{\gls{bc}}_{\gls{index},p}$) maintain a constant, predefined value, representing an external environment that acts as an infinite source or sink unaffected by internal processes. Conversely, compartment-derived boundary concentrations ($\gls{conc_vec}_{\gls{index}} $) continuously adapt to reflect the local internal state, allowing the boundary to evolve naturally with the model's internal dynamics.

\begin{equation}
\gls{bc} =
\sum_{\text{patch } p}
\begin{cases}
-  \gls{vol_flow_vec}^{\gls{bc}}_{p} \circ  \gls{conc_vec}^{\gls{bc}}_{\gls{index},p}   & \text{(fixed boundary)} \\
- \gls{vol_flow_vec}^{\gls{bc}}_{p} \circ \gls{conc_vec}_{\gls{index}}                    & \text{(zero-gradient / Neumann)}
\end{cases}
\label{eq:matrixFlowCalculationBC}
\end{equation}

\paragraph{Application to Multiphase Systems}
In multiphase flows the procedure above must be applied separately for each phase $\gls{phi_phase}$, with $\gls{Cell_label}$ denoting the cluster index to which a cell belongs. The phase-specific flow matrix is constructed by weighting the total volumetric flow rate at each face by the local phase volume fraction $\gls{vol_frac}^{\gls{phi_phase}}$:
\begin{equation}
\gls{vol_flow_mat}^{\gls{phi_phase}}_{j \rightarrow k} =
\begin{cases}
\gls{vol_frac}^{\gls{phi_phase}} \gls{vol_flow}_{j \rightarrow k} & \text{if } \gls{vol_flow} > 0 \text{ and } \gls{Cell_label}_{owner} \neq \gls{Cell_label}_{neighbor} \\
- \gls{vol_frac}^{\gls{phi_phase}} \gls{vol_flow}_{j \rightarrow k}  & \text{if } \gls{vol_flow} < 0 \text{ and } \gls{Cell_label}_{owner} \neq \gls{Cell_label}_{neighbor} \\
0 & \text{else}
\end{cases}
\label{eq:flowRateBetweenClusters}
\end{equation}

This weighting is not trivial in practice. It is strongly recommended to use the pre-integrated phase flow rate field directly from the \gls{cfd} solver (e.g.\ the $\gls{vol_frac}\gls{vol_flow}$ field available in \gls{of}), rather than interpolating the volume fraction separately to the faces. Naive face interpolation of $\gls{vol_frac}$ can produce incorrect phase volume fractions and, consequently, grossly incorrect phase flow rates, particularly near the gas--liquid interface. The same caution applies to boundary faces: the boundary flow rate must be multiplied by the face-centred phase volume fraction to obtain the correct phase-specific boundary flow rate.

\subsubsection{Mass Transfer}
Because the cluster boundaries are defined identically for all phases (\ref{sec:clustering}), each gas-phase compartment $i$ occupies the same spatial region as its liquid-phase counterpart. Interphase mass transfer therefore couples compartments of the same index across phases, with no cross-compartment exchange required. The phase volumes entering the balance are $\gls{Vol_vec}^{\gls{gas}} = \gls{Vol_vec}_{\gls{tot}} \circ \gls{vol_frac_vec}^{\gls{gas}}$ and $\gls{Vol_vec}^{\gls{liq}} = \gls{Vol_vec}_{\gls{tot}} \circ \gls{vol_frac_vec}^{\gls{liq}}$, respectively.

The concentration change due to interphase mass transfer in the gas and liquid phases is given by \eqref{eq:massTransferGas} and \eqref{eq:massTransferLiquid}:

\begin{equation}
\mathbf{s}_{\text{\gls{mass_trans}}}^{\gls{gas}} = - \gls{kLa_vec}_{i} \oslash \gls{vol_frac_vec}^{\gls{gas}} \circ  (\gls{conc_vec}^{\gls{gas},\gls{interface}}_{i}-\gls{conc_vec}_{i}^{\gls{liq}})
\label{eq:massTransferGas}
\end{equation}

\begin{equation}
\mathbf{s}_{\text{\gls{mass_trans}}}^{\gls{liq}}  =  \gls{kLa_vec}_{i} \oslash \gls{vol_frac_vec}^{\gls{liq}} \circ  (\gls{conc_vec}^{\gls{gas},\gls{interface}}_{i}-\gls{conc_vec}_{i}^{\gls{liq}})
\label{eq:massTransferLiquid}
\end{equation}

Here the vector $\gls{kLa_vec}_i$ contains the liquid-phase volumetric mass transfer coefficient referenced to the total compartment volume. We note in passing that in some applications (e.g., biotechnology) it is common to report the liquid-volume-referenced coefficient $\gls{kLa_liq} = \gls{kLa}_{i} / \gls{vol_frac}^{\gls{liq}}$ instead. Thus, such liquid-volume-referenced coefficients must be first converted. Finally, the equilibrium interface concentration is obtained via Henry's law: $\gls{conc_vec}^{\gls{gas},\gls{interface}}_{i} = \gls{henry_const}^*_i \circ \gls{conc_vec}_{i}^{\gls{gas}}$.

\subsection{CFD Model} \label{sec:cfd-simulation}
All \gls{cfd} simulations were carried out using the \texttt{reactingTwoPhaseEulerFoam} solver available in \gls{of}, extended by a minor modification to write the volumetric mass transfer coefficient (field \texttt{KD}) to disk at each output time. All governing equations can be found in literature \cite{omidi_integrated_2025}. In this Euler-Euler approach, each phase is treated as an interpenetrating continuum, solved separately, and coupled via exchange terms that depend on the the local volume fraction ($\gls{vol_frac}$). The latter is sometimes also referred to as local hold-up or phase fraction. Alongside this volumetric phase fraction, the solver computes a distinct velocity vector field for each individual phase. This allows the model to capture independent momentum changes governed by interphase forces such as drag and turbulent dispersion. Furthermore, to resolve the local chemical composition, the solver tracks the species mass fractions within each phase. The spatial distribution of these species is governed by convective transport alongside both molecular and turbulent diffusion.

\subsubsection{Simulation Setup}\label{sec:sim_setup}
The test case was adapted from the \texttt{bubbleColumnEvaporatingReacting} tutorial, this is a quasi-\gls{2d} geometry. Full simulation parameters are listed in \ref{app:simulation_parameters}.
 The system models the dissolution of oxygen ($O_2$) in water ($H_2O$) at \SI{20}{\celsius}. Mass transfer is active only in the bubble-flow regime; no mass transfer occurs in the droplet regime. Hence, we refer to this setup as the \glsentrylong{r2d2} (\gls{r2d2}).

Three dissolved oxygen-consuming reactions were considered in the liquid phase, mimicking microbial uptake:
\begin{itemize}
    \item A half-order reaction: $\mathrm{O_2}^{\gls{liq}} \xrightarrow{\gls{rrc05O}} \emptyset$, with rate $r = \gls{rrc05O}\,(c^{\gls{liq}}_{\mathrm{O_2}})^{0.5}$.
    \item A first-order reaction: $\mathrm{O_2}^{\gls{liq}} \xrightarrow{\gls{rrc1O}} \emptyset$, with rate $r = \gls{rrc1O}\,(c^{\gls{liq}}_{\mathrm{O_2}})^{1}$.
    \item A second-order reaction: $\mathrm{O_2}^{\gls{liq}} \xrightarrow{\gls{rrc2O}} \emptyset$, with rate $r = \gls{rrc2O}\,(c^{\gls{liq}}_{\mathrm{O_2}})^{2}$.
\end{itemize}
These reaction orders were chosen to span a range of nonlinearity and to yield a non-trivial steady state concentration distribution. The simulation setup is shown in  \autoref{fig:sim_geometry}:
\begin{figure}[H]
    \centering
    \includegraphics[width=0.5\columnwidth]{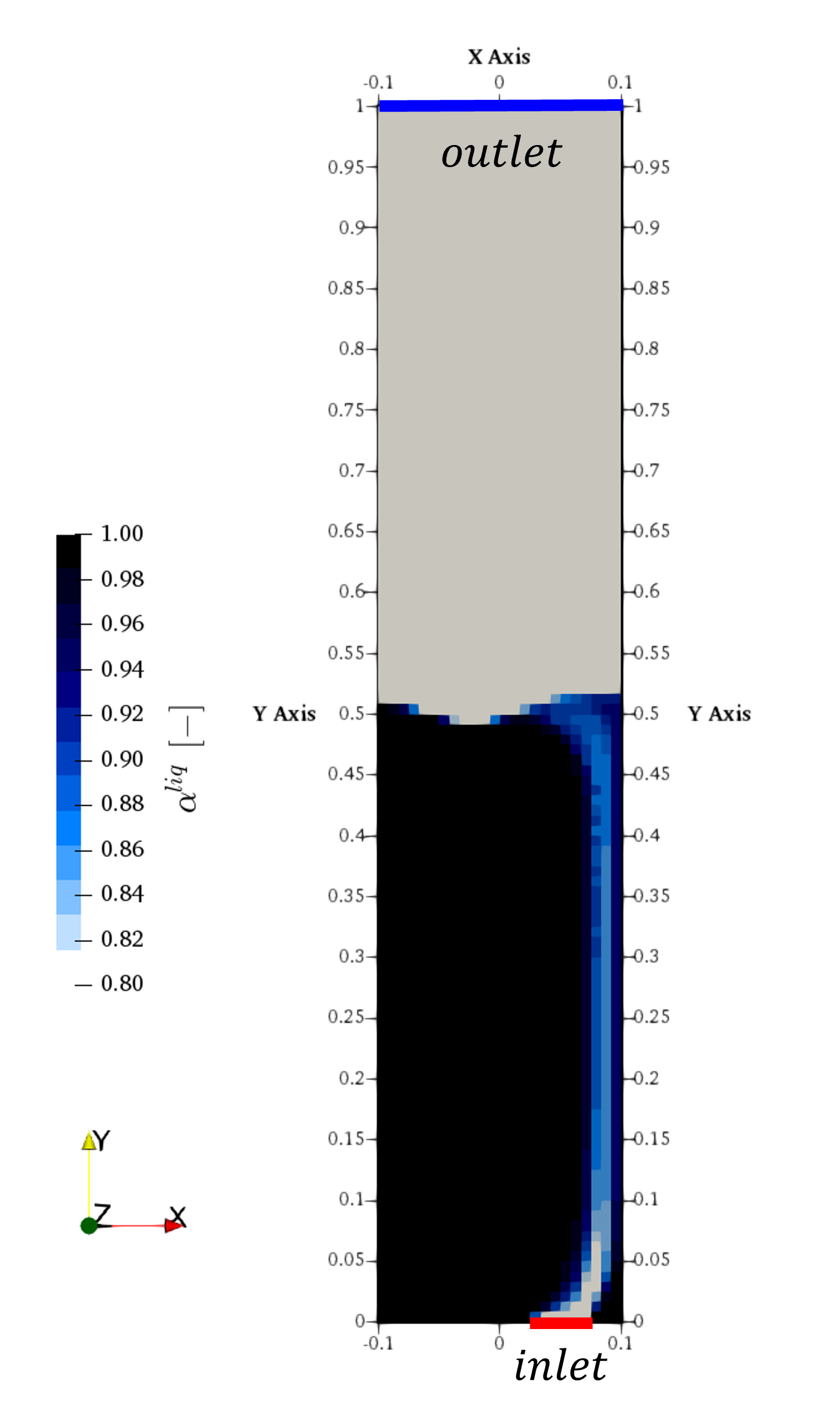}
    \caption{Simulation geometry with a quarter-width pure gas inlet on one side and initial filling height of \SI{0.5}{[\meter]}}
    \label{fig:sim_geometry}
\end{figure}
The initial liquid filling height is \SI{0.5}{\metre}. A mass transfer "blending" between a "bubbly" and a "droplet" flow regime is applied, which follows the local liquid volume fraction $\gls{vol_frac}^{\gls{liq}}$: below \num{0.4} the system is treated as fully in the droplet regime (no mass transfer), between \num{0.4} and \num{0.6} a linear blend is applied, above \num{0.6} the system is fully in the bubble regime with active mass transfer. The system level gas hold-up is computed as the mean gas volume fraction over all cells where $\gls{vol_frac}^{\gls{liq}} > 0.6$. ($\frac{1}{\gls{number}_{cells}} \sum \gls{vol_frac}^{\gls{gas}}_i$ exclusively for the $\gls{number}_{cells}$, where $\gls{vol_frac}^{\gls{liq}}_i > 0.6$ as all cells have the same size.)

\section{Numerical Methods} \label{sec:numerical_methods}

This section outlines the numerical framework developed to automatically generate and evaluate compartment models from time-averaged \gls{cfd} data. The workflow proceeds through the following steps:

\begin{enumerate}
    \item Clustering of the time-averaged \gls{cfd} data.
    \item Compute compartment model properties (e.g., flow, volume).
    \item Optimize flow rates to ensure mass conservation, if needed.
    \item Select mechanisms (e.g., reactions, mass transfer).
    \item Solve the resulting differential equations.
\end{enumerate}


\subsection{Clustering}\label{sec:clustering}

This section details the clustering algorithms employed in our framework. 
All clustering procedures were implemented using the Python library scikit-learn \cite{pedregosa_scikit-learn_2011}.

\subsubsection{K-Means Clustering with Graph Reassignment}

K-Means clustering is a partitioning method that divides a dataset into a user-defined number of distinct, non-overlapping groups. \cite{lloyd_least_1982}
Starting from randomly initialised cluster centres, the algorithm iteratively assigns each data point to the nearest centre and then recomputes each centre as the mean of its assigned points. This alternation continues until the assignments no longer change between iterations, at which point the algorithm has converged. The result is a partition that minimises the within-cluster spread in feature space. \cite{pedregosa_scikit-learn_2011}

Because k-means operates purely in feature space, the resulting clusters are not guaranteed to be spatially contiguous: cells carrying the same label may be geographically isolated from one another. For compartment modelling, every compartment must correspond to a connected region of the mesh. A graph-based post-processing step is therefore applied after clustering to enforce spatial connectivity. As this was already introduced by \cite{savarese_machine_2023} we will only shortly outline the strategy.

The \gls{cfd} mesh is first translated into a graph in which every computational cell is a node and every shared face between two neighbouring cells is an edge. After clustering, each node carries its cluster label. For every cluster, the subgraph formed by its cells is extracted and examined for connected components, i.e.\ isolated groups of cells that share no face with the rest of the cluster. Where such disconnected fragments exist, one of three actions is taken, and the scan is repeated until all clusters are fully connected:

\begin{enumerate}
    \item \textbf{Small fragments:} A disconnected fragment whose total volume falls below a set threshold relative to the domain volume is absorbed into the neighbouring cluster that shares the greatest number of faces with it. (In the present study, this limit was set to approximately 7 \gls{cfd} cells; however, the $k$-means algorithm consistently generated larger clusters, meaning this threshold was never reached in our cases.)
    \item \textbf{Single-cell fragments:} An isolated single cell is always absorbed into its most-connected neighbour, regardless of its volume.
    \item \textbf{Large fragments:} A disconnected fragment that exceeds the volume threshold is promoted to a new, independent cluster, thereby increasing the total cluster count.
\end{enumerate}

The neighbour selection in both absorption cases is determined by counting shared faces rather than comparing geometric distances, so the reassignment respects the actual mesh topology and preserves physical flow connectivity.

\subsubsection{Hierarchical (Agglomerative) Clustering}
Hierarchical clustering constructs a multilevel, nested organization of data. In this work, we employ an agglomerative (bottom-up) strategy, where individual data points are progressively merged into larger clusters based on a predefined linkage criterion \cite{murtagh_algorithms_2012}. These criteria govern the merging process: Ward's method minimizes within-cluster variance \cite{murtagh_wards_2014}, while complete, average, and single linkages compute inter-cluster distances based on maximum, mean, or minimum pairwise distances, respectively \cite{pedregosa_scikit-learn_2011}. Crucially for \gls{cfd} applications, this agglomerative approach allows the integration of a spatial connectivity matrix. By restricting the algorithm to only merge adjacent regions, this feature inherently guarantees the physical contiguity of the resulting compartments, making it a well suited method for spatial partitioning.

\subsection{Volume Flow Optimization}
As noted in \cite{savarese_machine_2023}, \gls{cfd} simulations are subject to numerical rounding errors, and the subsequent aggregation from cell-to-cell to compartment-to-compartment flows can introduce additional mass inconsistencies. In practice this means that more, or less, fluid may enter a compartment than leaves it, even in a steady-state simulation. The \gls{cm} therefore requires a dedicated verification and, if necessary, a correction step before the reactor network is used for simulation.

Physically, the goal of the correction is simple: adjust the inter-compartment flow rates by the smallest possible amount so that every compartment satisfies a steady-state mass balance, without ever creating a flow path between two compartments that are not spatially connected.

This section details a four-step volume flow optimization procedure designed to enforce steady-state mass conservation within the compartment model. The process begins with boundary condition aggregation, which separates the directional \gls{cfd} boundary flows into distinct inflow and outflow components for each compartment. Subsequently, a mass conservation check calculates both local compartment-level mismatches and global reactor-level imbalances, comparing them against a user-defined tolerance. If the verification fails, a two-stage correction is initiated. If there is a global inconstancy outlet correction is applied, by proportionally scaling all boundary outflows to restore the overall mass balance of the reactor. If local imbalances remain, a constrained optimization problem is solved via the \gls{slsqp} method, to minimally adjust inter-compartment flow rates, ensuring exact mass conservation without creating unphysical flow connections between spatially separated compartments.

    \paragraph{(i) Boundary Condition Aggregation}
    The \gls{cfd} flow from a boundary follows the \gls{of} sign convention: a positive value means fluid leaves the domain (outflow), a negative value means fluid enters (inflow). Boundary flows for each compartment $j$ are therefore separated by summing over all boundary patches $p$:
    \begin{equation}
        \gls{vol_flow_vec}^{\gls{bc}}_{\gls{in},j} = \sum_p \max\!\left(-\gls{vol_flow_vec}^{\gls{bc}}_{p,j},\, 0\right), \qquad
        \gls{vol_flow_vec}^{\gls{bc}}_{\gls{out},j} = \sum_p \max\!\left(\gls{vol_flow_vec}^{\gls{bc}}_{p,j},\, 0\right).
    \end{equation}

    \paragraph{(ii) Mass Conservation Check}
    For steady-state operation, the total volumetric flow entering each compartment $j$ must equal the flow leaving it. The inflow combines fluid arriving from neighbouring compartments (column $j$ of the flow matrix $\gls{vol_flow_mat}$), fluid entering through the reactor boundary, and any volumetric source term (can be positive or negative). The outflow combines fluid leaving to neighbours (row $j$ of $\gls{vol_flow_mat}$) and fluid exiting through the boundary:
    \begin{align}
        \gls{vol_flow_vec}^{\gls{local}}_{\gls{in},j}  &= \sum_{k} \gls{vol_flow_mat}_{kj} + \gls{vol_flow_vec}^{\gls{bc}}_{\gls{in},j}  + \gls{vol_flow_vec}_{\gls{source},j} \label{eq:local_inflow}  \\
        \gls{vol_flow_vec}^{\gls{local}}_{\gls{out},j} &= \sum_{k} \gls{vol_flow_mat}_{jk} + \gls{vol_flow_vec}^{\gls{bc}}_{\gls{out},j}. \label{eq:local_outflow}
    \end{align}
    The source term $\gls{vol_flow_vec}_{\gls{source},j}$ is only relevant when significant phase change occurs, e.g.\ for gas dissolving into liquid the disappearing gas volume is $\gls{vol_flow}^{\gls{gas}}_{\gls{source}} = -\gls{mol_flow}^{\gls{gas}} \gls{mol_weight}^{\gls{gas}} / \gls{density}^{\gls{gas}}$, with $\gls{mol_flow}^{\gls{gas}}$ estimated from the local $\gls{kLa}$ and concentration driving force.

    Two relative error metrics are computed and compared against a user-defined tolerance $\gls{rel_tol}$:

    Local imbalance: the relative flow mismatch for each individual compartment $j$:
    \begin{equation}
        \gls{error}^{\gls{local}}_{j} = \frac{\left|\gls{vol_flow_vec}^{\gls{local}}_{\gls{in},j} - \gls{vol_flow_vec}^{\gls{local}}_{\gls{out},j}\right|}{\max\!\left(\gls{vol_flow_vec}^{\gls{local}}_{\gls{in},j},\, \gls{vol_flow_vec}^{\gls{local}}_{\gls{out},j}\right)}
    \end{equation}

    Global imbalance: the relative mismatch between what enters and leaves the reactor as a whole (internal flows between compartments cancel and do not contribute):
    \begin{equation}
        \gls{error}^{\gls{global}} = \frac{\left|\gls{vol_flow_vec}_{\gls{in}}^{\gls{global}} - \gls{vol_flow_vec}_{\gls{out}}^{\gls{global}}\right|}{\max\!\left(\gls{vol_flow_vec}_{\gls{in}}^{\gls{global}},\, \gls{vol_flow_vec}_{\gls{out}}^{\gls{global}}\right)}
    \end{equation}
    where $\gls{vol_flow_vec}_{\gls{in}}^{\gls{global}} = \sum_j \gls{vol_flow_vec}^{\gls{bc}}_{\gls{in},j} + \sum_j \gls{vol_flow_vec}_{\gls{source},j}$ and $\gls{vol_flow_vec}_{\gls{out}}^{\gls{global}} = \sum_j \gls{vol_flow_vec}^{\gls{bc}}_{\gls{out},j}$.
    The \gls{cm} is accepted for simulation if $\gls{error}^{\gls{global}} < \gls{rel_tol}$ and $\gls{error}^{\gls{local}}_{j} < \gls{rel_tol}$ for all $j$.

    If a verification step fails, the flow rates are adjusted described in the following two stages.

    \paragraph{(iv) Outlet Correction.}
    First, the overall reactor mass balance is restored. If the total inflow to the reactor does not equal the total outflow, all outlet boundary flows are scaled by a common factor, analogous to adjusting all outlet valves by the same proportion, until global conservation holds:
    \begin{equation}
        \gls{vol_flow_vec}'^{\gls{bc}}_{\gls{out},j} = \gls{vol_flow_vec}^{\gls{bc}}_{\gls{out},j} \cdot \frac{\gls{vol_flow_vec}_{\gls{in}}^{\gls{global}}}{\gls{vol_flow_vec}_{\gls{out}}^{\gls{global}}}.
        \label{eq:bc_global_correction}
    \end{equation}
    After this step, the total inlet flow equals the total outlet flow for the network as a whole, but individual compartments may still be imbalanced.

    \paragraph{(iv) Local Flow Correction}
    To find flows $\bm{x}$ which produce physically meaningful results, i.e. no imbalanced flows, we formulate an optimization problem
    \begin{equation}
        \min_{\bm{x} \geq \bm{0}} \left\|\bm{x} - \bm{x}_0\right\|^2 \quad \text{subject to} \quad \bm{A} \bm{x} = \bm{b}.
        \label{eq:flow_optimization_qp}
    \end{equation}
    where the constraint that flows need to be larger or equal to zero is enforced by inequality.
    
    The original flow vector $\bm{x}_0$ is obtained by flattening the flow rate matrix $\bm{\Phi}$ and using only the strictly positive entries
    \begin{equation}
        \bm x_0 = \operatorname{vec}(\bm{\Phi})_{i \in \mathcal{I}}, \qquad \mathcal{I} = \{i | \operatorname{vec}(\bm{\Phi})_i > 0\}.
    \end{equation}
    This ensures that now new flow connection between compartments that were not originally connected is created.

    The elements of $\bm b$ contain the inflows subtracted from the corrected outflows
    \begin{equation}
        \bm b_j = \gls{vol_flow_vec}'^{\gls{bc}}_{\gls{out},j} - \gls{vol_flow_vec}^{\gls{bc}}_{\gls{in},j}-\gls{vol_flow_vec}_{\gls{source},j}.
    \end{equation}

    The network connectivity matrix $\bm{A}$ links outflows and inflows. 
    Each column corresponds to one flow connection $e = (j \to k)$, with $A_{ke} = +1$ (inflow to $k$) and $A_{je} = -1$ (outflow from $j$).

    The constrained optimization problem is solved via the \gls{slsqp} method.
    By design, the original network structure and zero self-flow ($\gls{vol_flow_mat}'_{jj} = 0$) is enforced.

\subsection{Error Calculation}

For representing the error of our compartment model we present the local concentration error $\gls{error}_{\gls{conc}}$. With 
$\gls{conc}_{map}$ being the compartment concentration mapped back on the \gls{cfd} grid. As the normalizing concentration $\gls{conc}_{norm}$ we choose the maximal concentration from the \gls{cfd} simulation.

\begin{equation}
    \gls{error}_{\gls{conc}} = \frac{
    \sqrt{
        \frac{1}{\sum\gls{Vol_vec}^{\gls{liq}}} \sum \left( \left( \gls{conc_vec}_{map} - \gls{conc_vec}_{\gls{cfd}}  \right)^2 \circ \gls{Vol_vec}^{\gls{liq}} \right)
      }
    }{
    \gls{conc}_{norm}
    }
    \label{eq:error_c_local}
\end{equation}

Additional error matrices can be found in the \ref{app:extended_results}.

\section{Results} \label{sec:results}

This section first presents the verification of the compartment model, followed by a detailed analysis of the a quarter-gassed bubble column, which is of particular interest due to its poor mixing behaviour. For the sake of brevity, we will  discuss the 0.5\textsuperscript{th} order reaction in more detail. This is since the 1\textsuperscript{st} and 2\textsuperscript{nd} order  cases  show very similar results (i.e., a very homogenous concentration distribution), and are hence of lower interest for compartment modelling. We note in passing that these results are test cases to demonstrate the model and the software, and are not meant to be representative of a specific real-world system. The main purpose of this section is to demonstrate the capabilities and limitations of the compartmentalization approach under controlled conditions, rather than to provide detailed insights into a particular reactor design or operating regime. Therefore, we deliberately chose simplified cases that isolate specific physical phenomena and allow for clear interpretation of the results, rather than attempting to capture the full complexity of a real reactor system.

\subsection{Verification against Analytical Solutions}
\begin{figure}[H]
\centering
    \includegraphics[width=0.5\columnwidth]{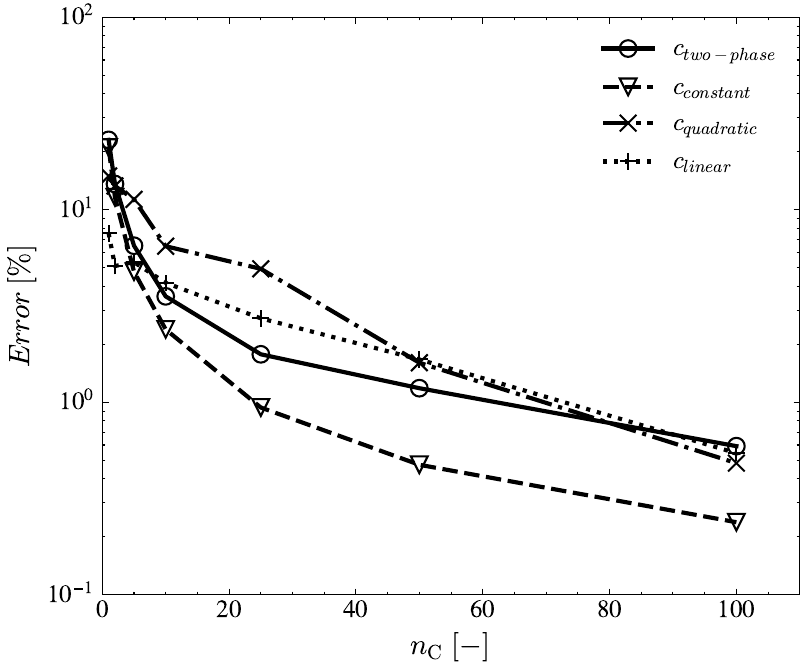}
    \caption[Compartment model verification] { Absolute error of the CM solution compared to the analytical solution for (i) reactive single phase (constant, linear , quadratic), and (i) non-reactive two-phase flow in a two-dimensional straight channel.}
    \label{fig:cm_vs_analyticalSolution}
\end{figure}

A detailed description of the setups and analytical derivations is provided in \ref{app:analytical_solutions}. The setups include three single phase scenarios (i.e., reactors with a constant, linear and quadratic velocity profile), as well as a two-phase scenario (i.e., a non-reactive case with constant velocity profile and mass transfer).
Our clustering approach utilizes local concentration and velocity profiles as baseline features. Because we assume no prior hierarchy among these variables, we do not apply feature weighting, meaning all features are treated with equal importance. To evaluate the model's accuracy, we measure the relative error of the total reactor yield for single-phase cases, and of the transferred species for the two-phase case. Since the feed is product-free, the total yield equals the exiting molar flow rate, and which is calculated as $\gls{mol_rate} = \sum \left( \gls{vol_flow_vec}^{\gls{bc}}_{\gls{out}} \circ \gls{conc}^{\gls{bc}}_{\gls{out}} \right)$. In the compartment model (\gls{cm}), the outlet concentration ($\gls{conc}^{\gls{bc}}_{\gls{out}}$) is approximated by the concentration of the final compartment. Because this final compartment averages fluid properties over its volume, it loses the precise spatial gradients present in the analytical reference. Therefore, the overall accuracy of the \gls{cm} is assessed using the relative error of the total yield: $\frac{\gls{mol_rate}_{analytical} - \gls{mol_rate}_{\gls{cm}}}{\gls{mol_rate}_{analytical}}$. The results are presented in \autoref{fig:cm_vs_analyticalSolution}, which illustrates this relative error across four different analytical concentration profiles ($c_{\text{two-phase}}$, $c_{\text{constant}}$, $c_{\text{linear}}$, and $c_{\text{quadratic}}$) as a function of the number of compartments ($\gls{number}_{\gls{cluster}}$). As expected, the error exhibits a sharp initial decline and decreases monotonically as the number of compartments increases. This consistent downward trend demonstrates that even when relying on unweighted features, the clustering algorithm successfully partitions the domain, steadily improving the macroscopic predictive accuracy as the compartmentalization becomes finer.

Our verification cases are intentionally conservative and serve as useful lower bounds on clustering performance. This is since real (reactor) systems feature diffusion phenomena and turbulent mixing mechanisms that are absent in our verification cases. Nevertheless, the monotonic error reduction demonstrates the robustness of our compartmentalization strategy for a variety of single- and two-phase scenarios.

\subsection{Bubble Column Reactor}
Details regarding the bubble column simulation setup can be found in Section \ref{sec:sim_setup}. As a first step, we characterize the baseline performance of the bubble column by determining the volumetric mass transfer coefficient, $\gls{kLa_liq}$. This parameter was extracted from a non-reacting CFD simulation using a virtual probe approach: nine probes were distributed throughout the liquid region, and the temporal evolution of the dissolved oxygen concentration at each probe was fitted to the standard exponential re-oxygenation model:
\begin{equation*}
\gls{conc}(\gls{time}) = \gls{conc}^{\gls{equil}} \left(1 - e^{-\gls{kLa_liq} \cdot \gls{time}}\right).
\end{equation*}
Both $\gls{kLa_liq}$ and the equilibrium concentration $\gls{conc}^{\gls{equil}}$ were fitted simultaneously for each probe; the system-level $\gls{kLa_liq}$ was then obtained by averaging the individual probe results. 

 To establish a baseline for the performance evaluation of our auto-generated compartment models, we report the predictive performance metrics of a conventional single-compartment model, denoted as "1-Comp (conv)". This baseline approach treats the entire reactor (in our case the liquid volume) as a single, perfectly mixed volume,  considers a single global mass balance for the liquid phase, and relies on the single volumetric mass transfer coefficient ($\gls{kLa_liq}$) described in the previous paragraph. Thereby, this formulation mirrors the standard lumped-parameter approach conventionally utilized in experimental reactor characterizations.

\subsubsection{Time Scale Analysis}\label{sec:time_scale_analysis}

Three time scales govern the interplay between transport and the moderately fast reactions considered in a gas-liquid reactor:
(i) The macro mixing time scale, $\gls{time_scale}_{\gls{mix}} = \gls{coord}/\gls{velo} = \gls{height}^{\gls{liq}}/{\gls{velo}^{\gls{in}}}$, characterises the convective transport of dissolved species over the reactor height. We note in passing that also a micro-mixing time scale (at the size of unresolved turbulent eddies) might be relevant for extremely fast reactions, which are not consider here.
(ii) The mass transfer time scale, $\gls{time_scale}_{\gls{mass_trans}} = 1/\gls{kLa_liq}$, represents the rate at which oxygen is transferred from the gas phase into the liquid.
(iii) The reaction time scale is given by $\gls{time_scale}_{\gls{react}} = \left(\gls{rrc}\,\gls{conc}^{\gls{react_order}-1}\right)^{-1}$, the unit of the reaction rate constant $\gls{rrc}$ varies accordingly with the reaction order $\gls{react_order}$.

To ensure a fair and physically meaningful comparison for different values of the reaction orders, the simulation cases were designed such that the characteristic time scales of mixing and reaction are matched. Across all cases, the operational and geometric parameters are kept constant, utilizing an inlet gas velocity $\gls{velo}_{\gls{in}}^{\gls{gas}} = \SI{5e-2}{\meter\per\second}$, a bubble diameter $\gls{diameter}^{\gls{gas}} = \SI{3e-3}{\meter}$, and an inital liquid height $\gls{height}^{\gls{liq}} = \SI{0.5}{\meter}$. Specifically, the reaction rate constants for each case were chosen such that all three cases yield identical mixing and reaction time scales ($\gls{time_scale}_{\gls{mix}} = \gls{time_scale}_{\gls{react}} = \SI{10}{\second}$). The mass transfer time scale is deliberately much longer ($\gls{time_scale}_{\gls{mass_trans}} = \SI{246}{\second}$) and is determined by the system's physical properties, yielding a system volumetric mass transfer coefficient of $\gls{kLa_liq} = \SI{4.059e-3}{\per\second}$. A summary of the varying case parameters is available in \autoref{tab:caseSetupTimeScales}.
Our design hence ensures that any differences in CLARA performance between the three cases originate solely from the nonlinearity of the kinetic expression (and the resulting concentration field), and are not based on differences in the underlying reaction time scale.

    \begin{table}[H]
        \centering
        \caption[Case setup with time scales]{Varying simulation case parameters and characteristic Hatta numbers. The reaction rate constant $\gls{rrc}$ was chosen for each reaction order $\gls{react_order}$ to yield equal mixing and reaction time scales.}
        \label{tab:caseSetupTimeScales}
        \begin{tabular}{ccc}
            \toprule
             $\gls{rrc}$ & $\gls{react_order}$   & $\gls{hatta_number}$\\
              & $[-]$ & $[-]$ \\
            \midrule
                0.1 & 1  & 0.15\\
                71.4 & 2  & 0.12\\
                3.74e-3 & 0.5  & 0.18\\
            \bottomrule
        \end{tabular}

    \end{table}

In all three cases, the Hatta number $\gls{hatta_number} < 1$ (see \autoref{eq:hatta_number}).
Therefore no enhancement of the mass transfer coefficient is expected \cite{danckwerts_absorption_1950}, i.e., mass transfer rates do not depend on the chemical reaction rate.

\begin{equation}
    \gls{hatta_number} = 
    \frac{
        \sqrt{
            \frac{2}{ \gls{react_order} +1 } \gls{rrc}{\gls{conc_vec}^{\gls{gas},\gls{interface}}}^{\gls{react_order}-1}\gls{diff_coeff}
        }
    }{
        \frac{\gls{kLa_liq} \gls{diameter}^{\gls{gas}}}{6 \gls{hold_up}}
    }
    \label{eq:hatta_number}
\end{equation}

\subsubsection{Flow Field}
\begin{figure}[H]
        \centering
        \begin{subfigure}[h]{0.24\columnwidth}
            \centering
            \includegraphics[width=\linewidth]{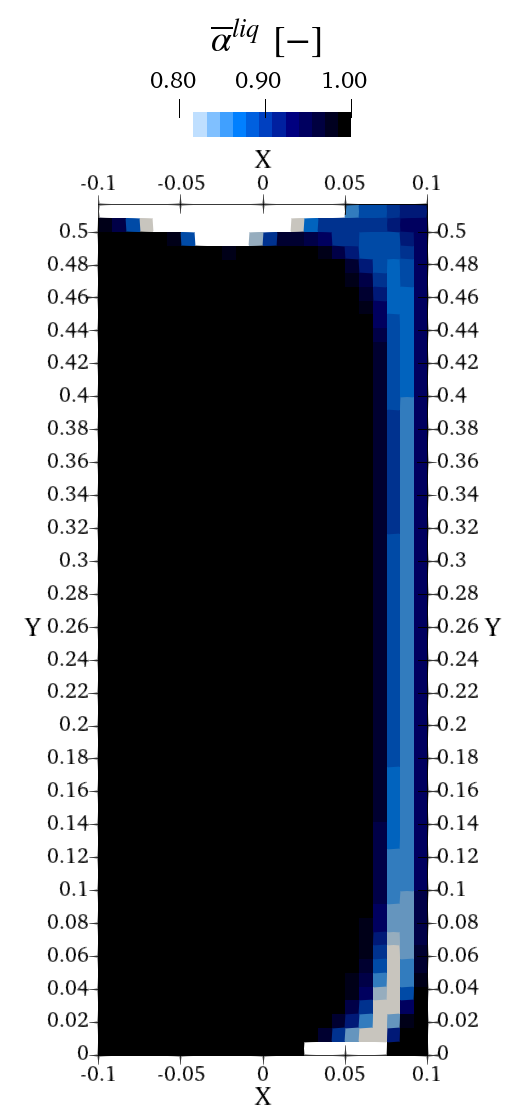}
            \caption{$\overline{\gls{vol_frac}}^{liq}$}
            \label{fig:t1810_alpha_liq}
        \end{subfigure}%
        \hfill %
        \begin{subfigure}[h]{0.24\columnwidth}
            \centering
            \includegraphics[width=\linewidth]{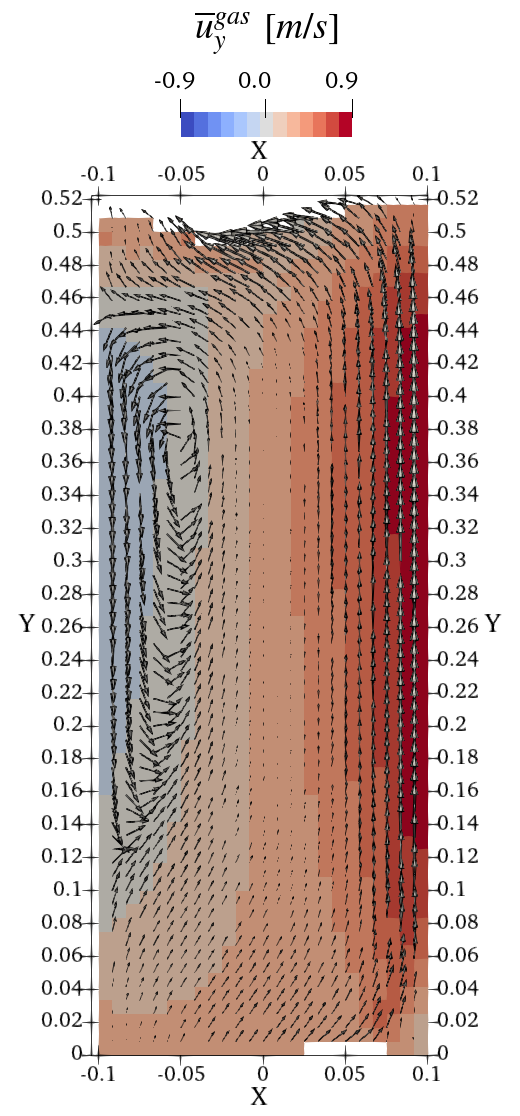}
            \caption{$\overline{\gls{velo}}^{gas}$}
            \label{fig:t1810_velo_gas_mean}
        \end{subfigure}%
        \hfill
        \begin{subfigure}[h]{0.24\columnwidth}
            \centering
            \includegraphics[width=\linewidth]{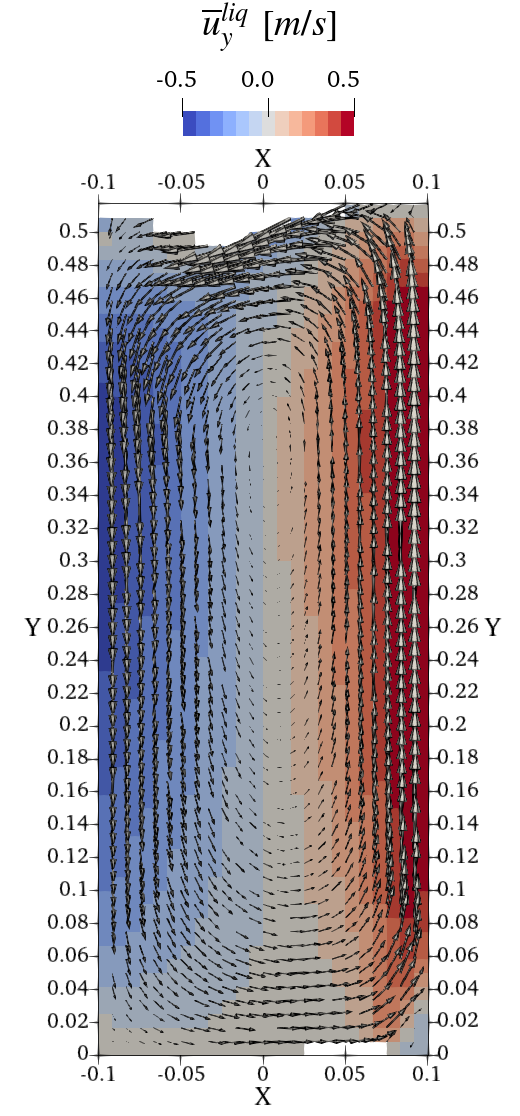}
            \caption{$\overline{\gls{velo}}^{liq}$}
            \label{fig:t1810_velo_liq_mean}
        \end{subfigure}%
        \hfill
        \begin{subfigure}[h]{0.24\columnwidth}
            \centering
            \includegraphics[width=\linewidth]{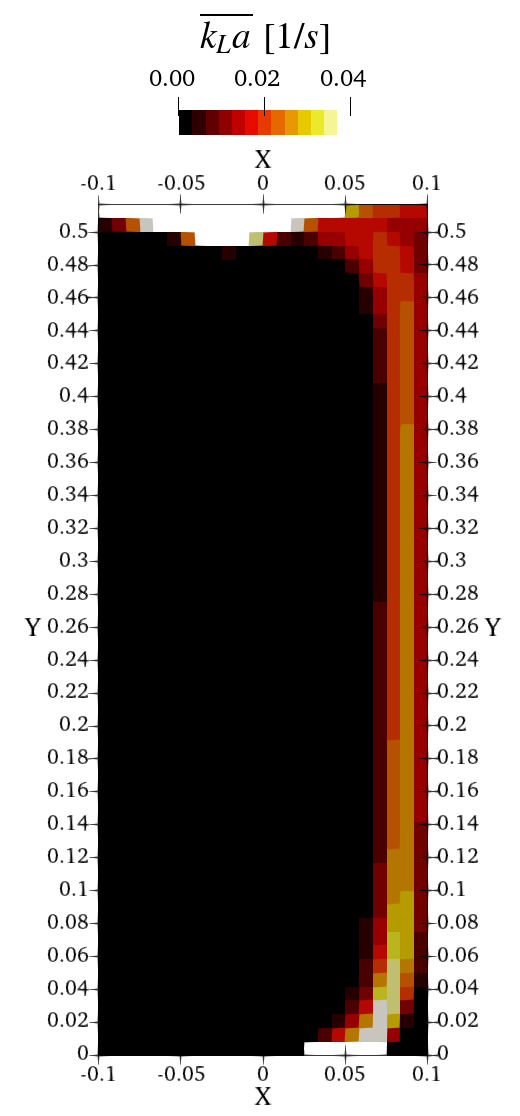}
            \caption{$\overline{\gls{kLa}}$}
            \label{fig:t1810_kLa_mean}
        \end{subfigure}   

        \caption[Time-averaged flow field of the quarter-gassed reactor]{Time-averaged flow field of the quarter-gassed reactor: (a) liquid volume fraction $\overline{\gls{vol_frac}}^{liq}$, (b) vertical gas velocity $\overline{\gls{velo}}^{gas}$, (c) vertical liquid velocity $\overline{\gls{velo}}^{liq}$, and (d) volumetric mass transfer coefficient $\overline{\gls{kLa}}$.}
        \label{fig:sim_results_t1810}
    \end{figure}

The time-averaged flow field for the quarter-gassed reactor is depicted in \autoref{fig:sim_results_t1810}. It is worth noting that, in this instance, the time-averaged values are essentially identical to the steady-state quantities.
Because gas is injected only through the right quarter of the bottom, the hydrodynamics are strongly asymmetric.
The global gas hold-up (i.e., \SI{2.16}{\percent}) is rather low, indicating that only a small fraction of the reactor volume contains dispersed bubbles.

The liquid volume fraction field (\autoref{fig:t1810_alpha_liq}) confirms this: $\overline{\gls{vol_frac}}^{liq}$ is close to unity throughout the domain, with the only deviation from unity being located at the lower-right region where the gas inlet is placed.
The gas velocity field (\autoref{fig:t1810_velo_gas_mean}) shows a pronounced upward plume on the gassed side, with the bubbles rising along the right wall and reaching velocities up to \SI{0.9}{\meter\per\second}.
The corresponding liquid velocity field (\autoref{fig:t1810_velo_liq_mean}) reveals a single large recirculation loop: the liquid is entrained upward by the rising bubbles on the right, and returns downward along the left side of the reactor, with magnitudes up to \SI{0.5}{\meter\per\second}.
The left-hand portion of the reactor therefore receives almost no gas-liquid contact, which is directly reflected in the local mass transfer coefficient distribution (\autoref{fig:t1810_kLa_mean}): $\overline{\gls{kLa}}$ is significant only in the gassed plume region and drops to near zero at the ungassed (i.e., left) side of the domain.

This spatially heterogeneous mass transfer distribution is the physical origin of the pronounced oxygen concentration gradients observed in the reacting cases \autoref{fig:reactionOrder_simResults}, and is precisely the type of structure that a compartment model must resolve in order to be predictive.

\subsubsection{Concentration Distribution}
\begin{figure}[H]
        \centering
        \begin{subfigure}[b]{0.3\columnwidth}
            \centering
            \includegraphics[width=\linewidth]{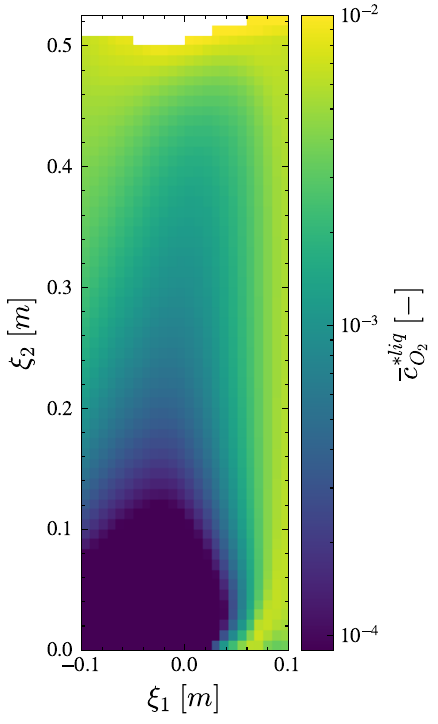} 
            \caption{0.5\textsuperscript{th} Order}
            \label{fig:t1871_cO2liq_simResult}
        \end{subfigure}
        \begin{subfigure}[b]{0.3\columnwidth}
            \centering
            \includegraphics[width=\linewidth]{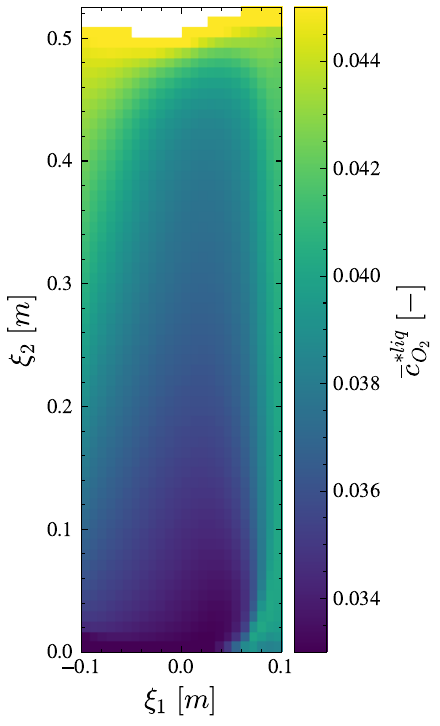}
            \caption{1\textsuperscript{st} Order}
            \label{fig:t1671_cO2liq_simResult}
        \end{subfigure}
        \begin{subfigure}[b]{0.3\columnwidth}
            \centering
            \includegraphics[width=\linewidth]{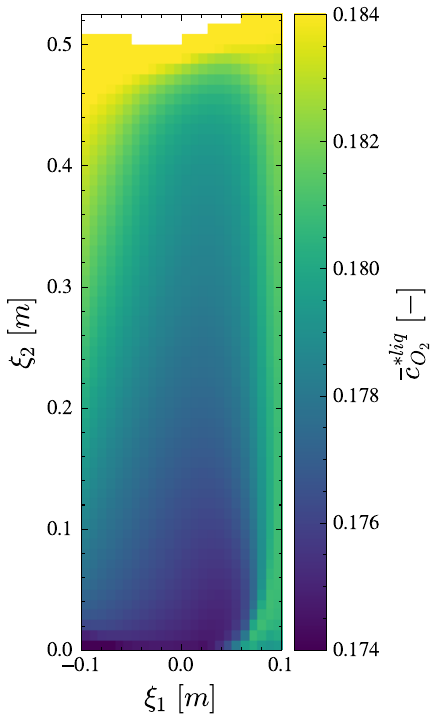}
            \caption{2\textsuperscript{nd} Order}
            \label{fig:t1771_cO2liq_simResult}
        \end{subfigure}

        \caption[Steady-state liquid-phase oxygen concentration for different reaction orders]{Steady-state normalised liquid-phase oxygen concentration $\overline{\gls{conc}}^{*,\gls{liq}}_{O_2}$ obtained from the CFD reference simulation for 0.5\textsuperscript{th}-, 1\textsuperscript{st}-, and 2\textsuperscript{nd}-order kinetics. The 1\textsuperscript{st}- and 2\textsuperscript{nd}-order feature a linear colorbar, while the 0.5\textsuperscript{th}-order case features a logarithmic colour bar.}
        \label{fig:reactionOrder_simResults}
    \end{figure}

Having established the hydrodynamic flow field, we now examine the time averaged (steady-state) oxygen concentration profiles obtained from the full CFD simulations for the three reaction orders under consideration.
\autoref{fig:reactionOrder_simResults} compares the normalised liquid-phase oxygen concentration for 1\textsuperscript{st}-, 2\textsuperscript{nd}-, and 0.5\textsuperscript{th}-order kinetics.
For the 1\textsuperscript{st}- and 2\textsuperscript{nd}-order cases, the concentration field is nearly spatially uniform
(increasing the reaction rate constant by a factor of 10 and 100 still resulted in fairly homogeneous concentration distributions, data not shown).
In contrast, the 0.5\textsuperscript{th}-order case exhibits a pronounced spatial gradient spanning several orders of magnitude (note the logarithmic colour scale in \autoref{fig:t1871_cO2liq_simResult}), with the lowest concentrations appearing near the bottom left region of the reactor.
This strong spatial heterogeneity is due to the sub-linear reaction order, which sustains a finite reaction rate even in low concentration  regions, which ultimately become depleted of dissolved oxygen. This makes this case the most demanding benchmark for a compartment model.

\begin{figure}[H]
    \centering
        \includegraphics[width=0.7\columnwidth]{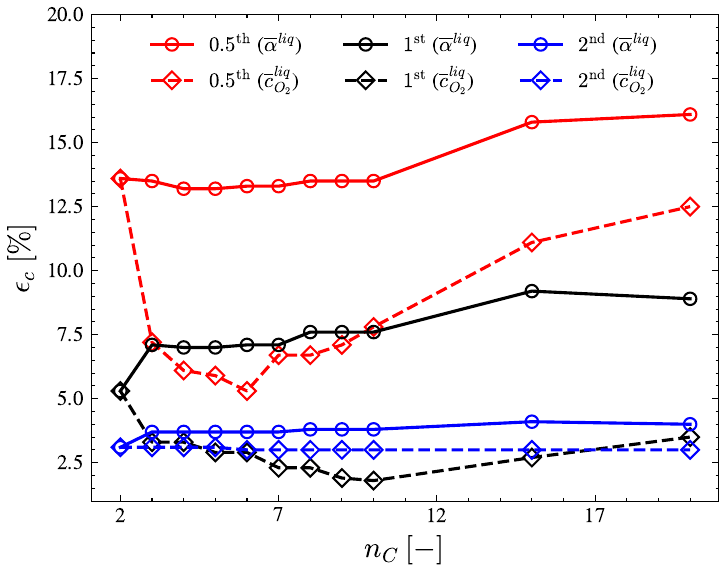}
        \caption[Local concentration error for different reaction orders using agglomerative clustering]{Local concentration error $\epsilon_c$ as a function of the number of clusters $\gls{number}_{\gls{cluster}}$ used during clustering. For the three reaction orders, obtained with agglomerative clustering using obvious features. Solid lines: volume fraction $\overline{\alpha}^{liq}$ as clustering feature; dashed lines: dissolved oxygen concentration $\overline{\gls{conc}}^{liq}_{O_2}$ as feature.}
        \label{fig:comarison_error_local_c_agglomerative}
    \end{figure}

The predictive performance for compartment models build with  CLARA are quantified in \autoref{fig:comarison_error_local_c_agglomerative}. This figure shows the (mean) local concentration error $\epsilon_c$ as a function of (i) the number of clusters $\gls{number}_{\gls{cluster}}$ and (ii) for different features (the agglomerative clustering algorithm was used in all cases). with obvious flow features.
    For the 1\textsuperscript{st}- and 2\textsuperscript{nd}-order reactions, the errors remain small across the entire range of $\gls{number}_{\gls{cluster}}$ (i.e., below \SI{10}{\percent} and \SI{5}{\percent}, respectively). This is consistent with the almost homogeneous concentration distribution (see \autoref{fig:reactionOrder_simResults}). Also, using the local dissolved oxygen concentration as the feature for clustering offers a moderate (for the 2\textsuperscript{nd}-order reaction) or high (for the 1\textsuperscript{st}-order reaction) advantage in terms of the error $\epsilon_c$.
    The 0.5\textsuperscript{th}-order case, however, yields substantially higher errors for both features (with the local dissolved oxygen concentration as the feature offering a substantial advantage for most values of $\gls{number}_{\gls{cluster}}$). Interestingly, and counter-intuitively, the error \emph{increases} with increasing $\gls{number}_{\gls{cluster}}$ for $\gls{number}_{\gls{cluster}}> 7$ in case of the 0.5\textsuperscript{th}-order reaction case.
    This unexpected trend suggests that the default choice of clustering features is poorly suited to resolve the sharp concentration gradients present in the 0.5\textsuperscript{th}-order case, motivating our deeper investigations in the following subsections.

\subsubsection{Details for the 0.5th-order reaction}\label{sec:t1871_results}
As established in the previous subsection, the 0.5\textsuperscript{th}-order case is the most demanding test for \gls{clara} due to its strongly heterogeneous concentration field.
We therefore examine this case in detail, focusing on the influence of the feature selection on the compartment models' predictive power.

A practically important observation concerns the effective mass transfer coefficient in the reacting simulation: the spatially averaged value obtained from the \gls{cfd} solution is $\gls{kLa_liq} = \SI{3.967e-3}{[1/\second]}$, which is just slightly (i.e., -2.27\%) below the \SI{4.059e-3}{[1/\second]} fitted from the non-reacting case (see \autoref{tab:caseSetupTimeScales}).
This small reduction is consistent with a modest decrease in gas hold-up, which is caused by bubble shrinkage as oxygen is consumed by the reaction: At a fixed bubble size (which was the case in our simulation setup), the reduction in gas volume fraction directly reduces the interfacial area available for mass transfer.

\begin{figure}[H]
        \centering
        \begin{subfigure}[b]{0.3\columnwidth}
            \centering
            \includegraphics[width=\linewidth]{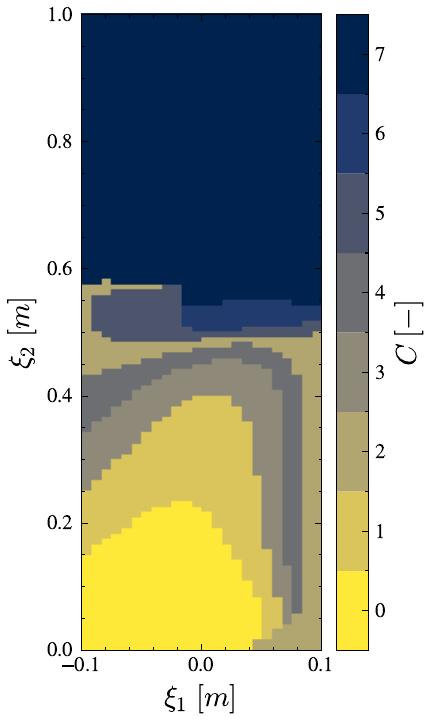}
            \caption{Cluster index}
            \label{fig:t1871_clara_O2liq_C8}
        \end{subfigure}
        \begin{subfigure}[b]{0.3\columnwidth}
            \centering
            \includegraphics[width=\linewidth]{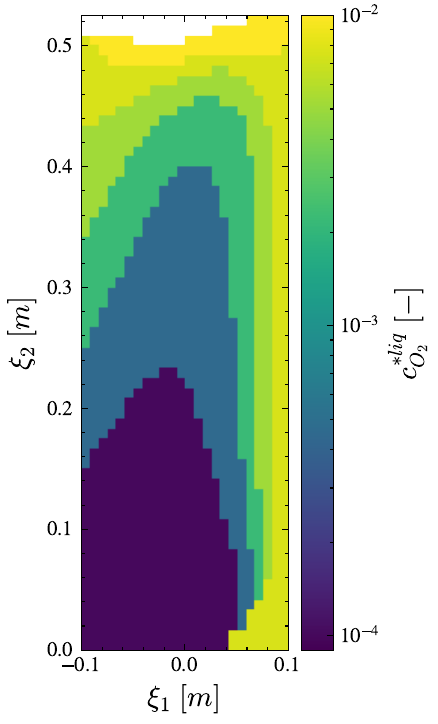}
            \caption{CLARA}
            \label{fig:t1871_c_O2_liquid_R2D2_fromCLARA}
        \end{subfigure}  
        \begin{subfigure}[b]{0.3\columnwidth}
            \centering
            \includegraphics[width=\linewidth]{figures/t1871_cStar_O2_liquid_R2D2_fromOF.pdf}
            \caption{OpenFoam}
            \label{fig:t1871_c_O2_liquid_R2D2_fromOF}
        \end{subfigure}
        \caption[CLARA clustering and concentration results for the 0.5\textsuperscript{th}-order case]{Representative \gls{clara} solution for the 0.5\textsuperscript{th}-order reaction using $\gls{number}_{\gls{cluster}} = 8$ clusters (i.e., 16 compartments) with $\overline{\gls{conc}}^{\gls{liq}}_{O_2}$ as the clustering feature. (a) Cluster index: a single large cluster captures the headspace region, while seven clusters resolve the concentration gradients in the active part of the reactor (the domain is shrunk in the vertical direction). (b) \gls{clara} concentration prediction and (c) \gls{cfd} reference CFD result.}
        \label{fig:t1871_results}
    \end{figure}

\autoref{fig:t1871_results} illustrates a representative \gls{clara} solution for this case. The solution was obtained using the time-averaged dissolved oxygen concentration $\overline{\gls{conc}}^{\gls{liq}}_{O_2}$ as the clustering feature with $\gls{number}_{\gls{cluster}} = 8$ clusters, which yields 16 individual compartments, 8 for each phase according to the volume fraction.
Panel \ref{fig:t1871_clara_O2liq_C8} shows the resulting cluster indices.
The feature choice is physically well-motivated: because $\overline{\gls{conc}}^{\gls{liq}}_{O_2}$ directly reflects the spatial distribution of the quantity the compartment model must predict, the algorithm partitions the domain according to the concentration field.
Consequently, the upper \SI{40}{[\percent]} of the reactor is assigned to a single large cluster, while the remaining seven clusters partition the lower region where the concentration gradient is high.
The lowest-concentration cluster (shown in yellow, and with cluster index $C = 0$) correctly identifies the oxygen-depleted zone at the bottom left of the bubble column.

Panels (\ref{fig:t1871_c_O2_liquid_R2D2_fromCLARA}) and (\ref{fig:t1871_c_O2_liquid_R2D2_fromOF}) compare the resulting \gls{clara} concentration field with the \gls{cfd} reference, both on a logarithmic scale.
The compartment model reproduces the dominant spatial features of the reference field: the strongly depleted back flow region in the lower-left part of the reactor, an  intermediate zone with strong concentration gradients, and the high concentration region close to the plume and the gas-liquid interface.
The staircase-like appearance of the \gls{clara} field is a natural consequence of the discrete compartment representation which is cell-based.
    
\begin{figure}[H]
        \centering
        \begin{subfigure}[h]{0.48\columnwidth}
            \centering
            \includegraphics[width=\linewidth]{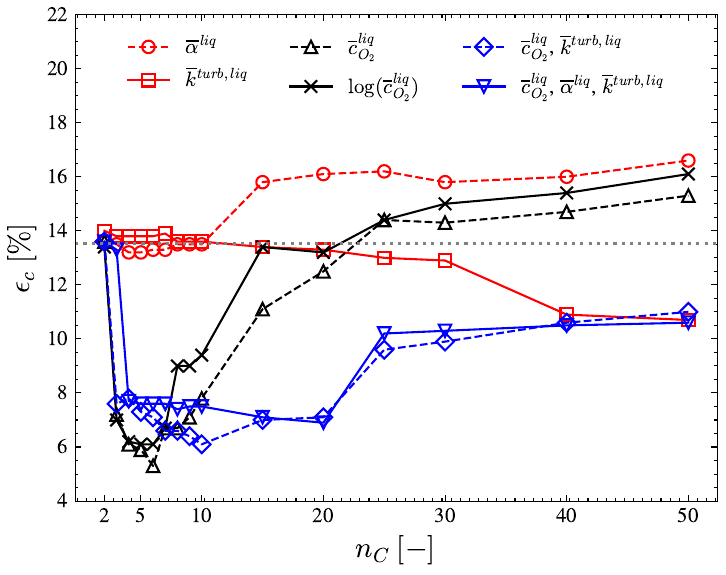}
            \caption{Agglomerative clustering}
            \label{fig:t1871_error_agglo}
        \end{subfigure}%
        \hfill
        \begin{subfigure}[h]{0.48\columnwidth}
            \centering
            \includegraphics[width=\linewidth]{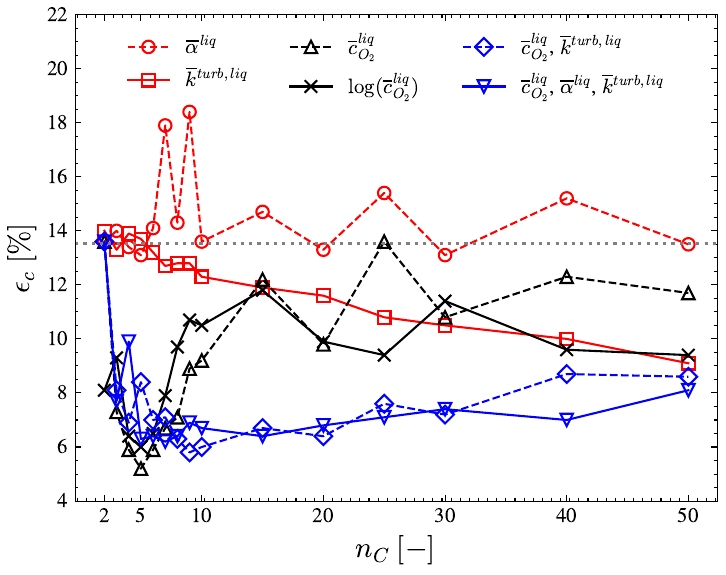}
            \caption{K-means clustering}
            \label{fig:t1871_error_kmeans}
        \end{subfigure}%

        \caption[Feature sensitivity of $\epsilon_c$ for the 0.5\textsuperscript{th}-order case]{Local concentration error $\epsilon_c$ \eqref{eq:error_c_local} for the 0.5\textsuperscript{th}-order case as a function of the number of clusters $\gls{number}_{\gls{cluster}}$ for six feature combinations, comparing agglomerative and k-means clustering. The dotted gray line
    (\tikz[baseline=-0.5ex]\draw[gray, dotted, line width=1.0pt] (0,0) -- (0.3,0);)
    marks the error of the conventional single-compartment model.}
        \label{fig:t1871_error}
    \end{figure}

To assess this behaviour systematically, \autoref{fig:t1871_error} compares the local concentration error $\epsilon_c$ as a function of $\gls{number}_{\gls{cluster}}$ for both agglomerative and k-means clustering, across different clustering features.

Several important trends emerge from \autoref{fig:t1871_error}:
first, purely flow-based features prove insufficient for this case.
Using $\overline{\gls{vol_frac}}^{\gls{liq}}$ alone yields errors that exceed the single-compartment baseline across the entire range of $\gls{number}_{\gls{cluster}}$ for agglomerative clustering, while the turbulent kinetic energy $\overline{\gls{tke}}^{\gls{liq}}$ alone only marginally improves upon it at large $\gls{number}_{\gls{cluster}}$.
Both features are insensitive to the concentration field, and therefore cannot guide the algorithm to partition the reactor in a meaningful way.

Second, concentration-based features (shown as black symbols in \autoref{fig:t1871_error}) achieve a pronounced error minimum at small compartment numbers,  but then the error increases with increasing $\gls{number}_{\gls{cluster}}$.
Using the dissolved oxygen feature $\overline{\gls{conc}}^{\gls{liq}}_{O_2}$ attains the lowest error of the entire study (i.e., $\epsilon_c \approx \SI{4.5}{[\percent]}$ at $\gls{number}_{\gls{cluster}} = 5$ with agglomerative clustering), well below the single-compartment baseline of \SI{13.5}{[\percent]}.
Beyond this optimum, however, the error increases monotonically with $\gls{number}_{\gls{cluster}}$, eventually surpassing the baseline.
A qualitatively identical but less pronounced trend is observed for $\log(\overline{\gls{conc}}^{\gls{liq}}_{O_2})$.

Third, multi-feature combinations provide the most robust performance.
Combining concentration-based features with the turbulent kinetic energy feature does not exhibit such a strong increasing trend as seen for single variable features.
Agglomerative clustering produces smoother, more monotone error curves compared to k-means, which may be attributed to the graph reassignment algorithm applied as a post-processing step rather than being built into the clustering itself.

Next, we aim at investigating the root cause of the counter-intuitive error increase with $\gls{number}_{\gls{cluster}}$. Thereby our idea is to isolate the effect of a phenomenon not included in our compartment model: turbulent mixing.

\subsubsection{Effects due to Turbulent Mixing}
Three distinct mechanisms drive the counter-intuitive error increase observed at higher $\gls{number}_{\gls{cluster}}$,  and hence reveal fundamental limitations of a purely concentration-based feature selection.

\begin{figure}[H]
        \centering
        \begin{subfigure}[b]{0.29\columnwidth}
            \centering
            \includegraphics[width=\linewidth]{figures/t1871_cStar_O2_liquid_R2D2_fromOF.pdf}
            \caption{\gls{cfd} reference}
            \label{fig:t1871_c_O2_liquid_R2D2_fromOF_app}
        \end{subfigure}
        \hfill
        \begin{subfigure}[b]{0.29\columnwidth}
            \centering
            \includegraphics[width=\linewidth]{figures/t1871_c_O2_liquid_R2D2_fromCLARA_C8.pdf}
            \caption{\gls{clara} ($\gls{number}_{\gls{cluster}} = 8$)}
            \label{fig:t1871_c_O2_liquid_R2D2_fromCLARA_C8}
        \end{subfigure}
        \hfill
        \begin{subfigure}[b]{0.29\columnwidth}
            \centering
            \includegraphics[width=\linewidth]{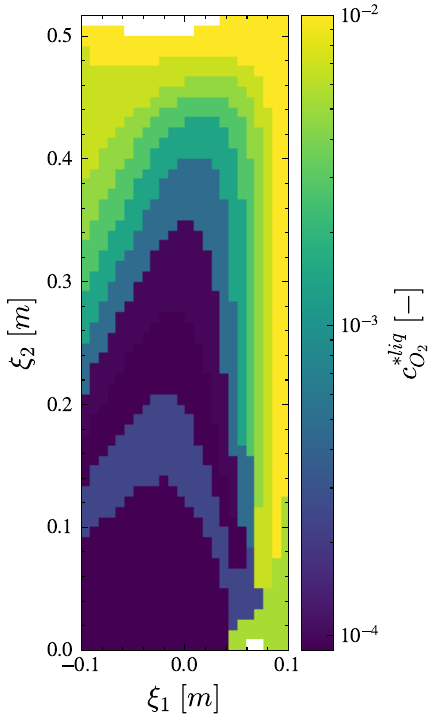}
            \caption{\gls{clara} ($\gls{number}_{\gls{cluster}} = 16$)}
            \label{fig:t1871_c_O2_liquid_R2D2_fromCLARA_C16}
        \end{subfigure}

        \caption[Effect of over-clustering on the predicted concentration field]{Effect of over-clustering on the predicted concentration field for the 0.5\textsuperscript{th}-order reaction, with $\overline{\gls{conc}}^{\gls{liq}}_{O_2}$ as the used feature. Increasing $\gls{number}_{\gls{cluster}}$ from 8 to 16 enlarges the predicted depletion zone (purple color) well beyond the \gls{cfd} reference, a direct consequence of compartments bridging the bubble-rich and the return-flow region.}
        \label{fig:t1871_OF_clara_8vs16}
    \end{figure}

The first mechanism is geometric fragmentation:
as $\gls{number}_{\gls{cluster}}$ increases, a concentration-based feature causes the algorithm to partition the domain into thin, contoured bands that follow the iso-concentration surfaces.
These regions have extreme surface-area-to-volume ratios. Consequently, the fluid residence time within each zone may become much shorter than the characteristic mixing time \cite{tajsoleiman_cfd_2019}.
More critically, these thin bands span laterally across the reactor, connecting the bubble-rich plume on the right (where oxygen is transferred from the gas to the liquid phase), with the bubble-free descending return flow on the left (where oxygen is consumed by the reaction with no replenishment).
By averaging the volumetric mass transfer coefficient $\gls{kLa_liq}$ across both regions, the compartment model artificially applies gas-phase oxygen supply to zones that should be oxygen depleted, thereby significantly overestimating the local concentration.
This effect is clearly visible in \autoref{fig:t1871_OF_clara_8vs16}: while the 8-cluster solution reproduces the general shape of the reference concentration field (see panel a), the 16-cluster solution shows a markedly enlarged depletion zone.

The second mechanism is the suppression of (turbulent) mixing at compartment boundaries when picking large values for $\gls{number}_{\gls{cluster}}$:
as the compartment resolution increases, compartment boundaries inevitably bisect well mixed zones.
Because the inter-compartmental exchange in the present model is governed strictly by convective flow rates (taken from the time-average velocity field), it does not capture  turbulent mixing rates that may be present at compartment boundaries. The stabilising effect of including $\overline{\gls{tke}}^{\gls{liq}}$ as a clustering feature, observed in \autoref{fig:t1871_error}, supports this interpretation: by guiding compartment boundaries away from  zones with high turbulently intensity, a feature set involving $\overline{\gls{tke}}^{\gls{liq}}$ indirectly reduces the bisection of well mixed zones.

\begin{figure}[H]
        \centering
        \begin{subfigure}[b]{0.3\columnwidth}
            \centering
            \includegraphics[width=\linewidth]{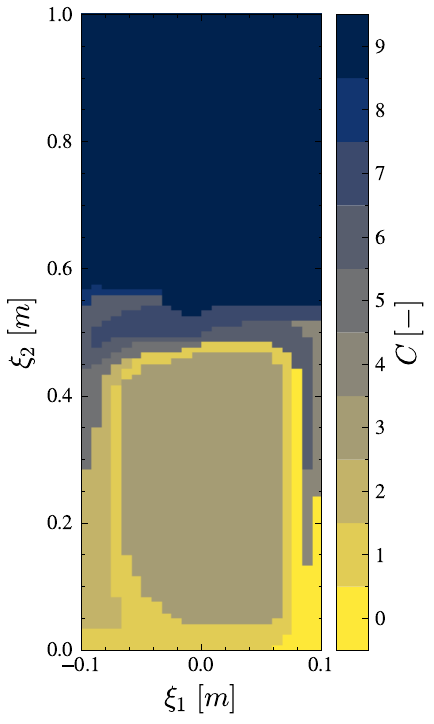}
            \caption{Cluster index}
            \label{fig:t1875_clara_logwO2liq_C10}
        \end{subfigure}
        \begin{subfigure}[b]{0.3\columnwidth}
            \centering
            \includegraphics[width=\linewidth]{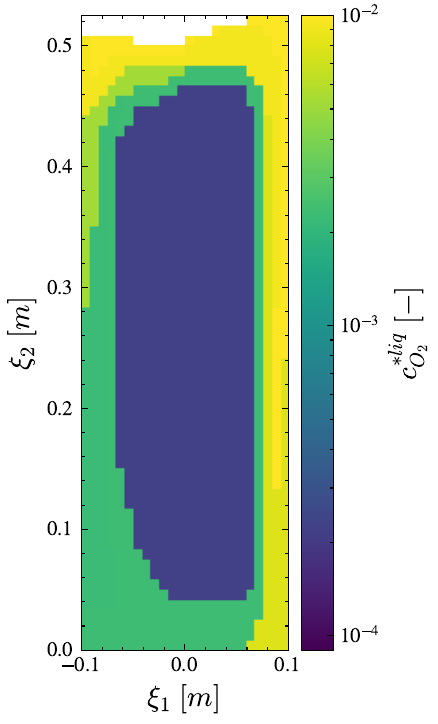}
            \caption{\gls{clara} ($\gls{number}_{\gls{cluster}} = 10$)}
            \label{fig:t1875_c_O2_liquid_R2D2_fromCLARA}
        \end{subfigure}
        \begin{subfigure}[b]{0.3\columnwidth}
            \centering
            \includegraphics[width=\linewidth]{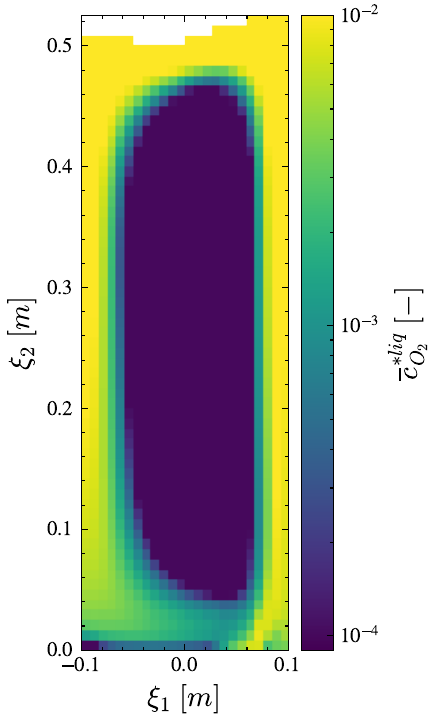}
            \caption{\gls{cfd} reference}
            \label{fig:t1875_c_O2_liquid_R2D2_fromOF}
        \end{subfigure}
        \caption[CLARA clustering and concentration results for the 0.5\textsuperscript{th}-order case with high $\gls{schmidt_number_turbulent}$ ]{Concentration field for the 0.5\textsuperscript{th}-order reaction with $\gls{schmidt_number_turbulent} = \num{1e7}$, obtained with agglomerative clustering using $\overline{\gls{conc}}^{\gls{liq}}_{O_2}$ and $\gls{number}_{\gls{cluster}} = 10$ clusters. (a) Cluster index:  A single large cluster captures the headspace region, suppressing turbulent diffusivity creates a large depleted bulk region surrounded by a thin saturation rim, within the liquid region. (b) \gls{clara} concentration prediction and (c) \gls{cfd} reference CFD result.}
        \label{fig:t1875_results}
    \end{figure}

To test our hypothesis that turbulent mixing is the dominant mixing mechanism in our bubble column reactor, a CFD simulation was performed in which the turbulent Schmidt number was raised to $\gls{schmidt_number_turbulent} = \num{1e7}$. This effectively suppresses the turbulent mixing rate, while leaving molecular diffusion unchanged. The resulting concentration field and a representative \gls{clara} solution are shown in \autoref{fig:t1875_results}.

The effect on the CFD reference concentration field is dramatic: with turbulent diffusivity suppressed, the dissolved oxygen concentration profile narrows to a thin ring-like structure spanning along the outer edge of the domain. The resulting concentration field is markedly more heterogeneous than in the baseline case, with a large oval-shaped depleted zone.

Our clustering algorithm correctly identifies this structure: the cluster with index $C = 4$ forms a compact central region that maps onto the oxygen-depleted zone in the CFD simulation, while the remaining clusters resolve the large concentration gradients at the periphery. \gls{clara} reproduces the reference field well: the extent and shape of the depletion zone in panel (b) closely matches the \gls{cfd} result in panel (c). This confirms that, when the concentration field is dominated by reactions and convective transport (rather than by turbulent mixing), the compartment model is fully capable of capturing the spatial heterogeneity. Consequently, the degraded performance observed in the baseline case (see \autoref{fig:t1871_error} and the associated discussion) is indeed attributed to the elimination of turbulent mixing at compartment boundaries.

\begin{figure}[H]
        \centering
        \begin{subfigure}[h]{0.48\columnwidth}
            \centering
            \includegraphics[width=\linewidth]{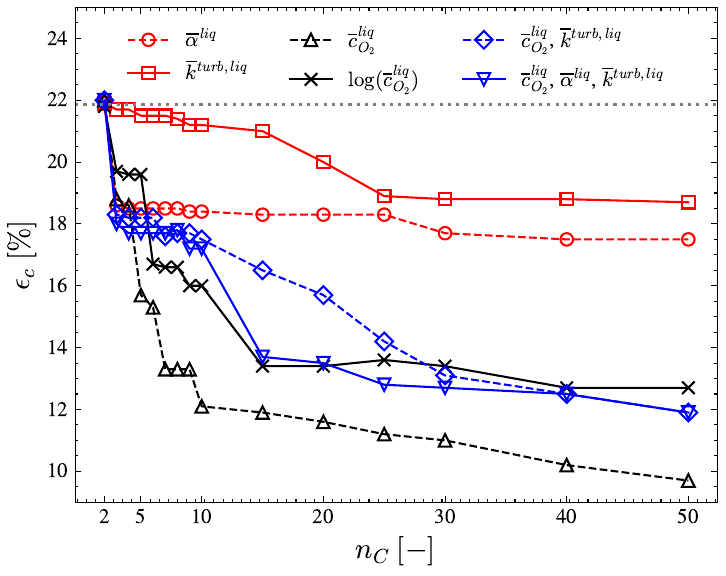}
            \caption{Agglomerative clustering}
            \label{fig:t1875_error_agglo}
        \end{subfigure}%
        \hfill
        \begin{subfigure}[h]{0.48\columnwidth}
            \centering
            \includegraphics[width=\linewidth]{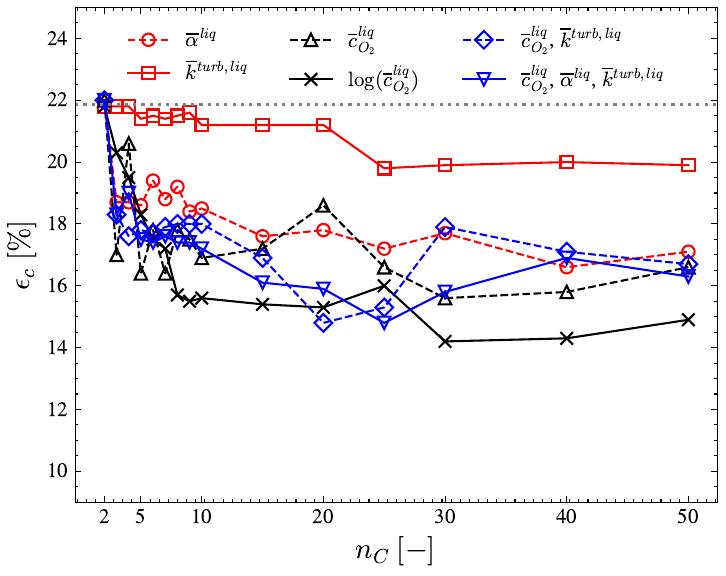}
            \caption{K-means clustering}
            \label{fig:t1875_error_kmeans}
        \end{subfigure}%

        \caption[Feature sensitivity of $\epsilon_c$ with suppressed turbulent diffusivity ($\gls{schmidt_number_turbulent} = \num{1e7}$)]{Local concentration error $\epsilon_c$ \eqref{eq:error_c_local} as a function of $\gls{number}_{\gls{cluster}}$ for the case with suppressed turbulent diffusivity ($\gls{schmidt_number_turbulent} = \num{1e7}$). The dotted gray line
    (\tikz[baseline=-0.5ex]\draw[gray, dotted, line width=1.0pt] (0,0) -- (0.3,0);)
    marks the single-compartment baseline (\SI{22}{[\percent]}).}
        \label{fig:t1875_error}
    \end{figure}

The most important result in \autoref{fig:t1875_error} is that for the case without turbulent mixing (in the CFD simulation) the compartment models shows reduced errors upon an increasing of $\gls{number}_{\gls{cluster}}$. Specifically, all concentration-based feature sets now show a monotonically decreasing error with increasing number of clusters, reaching approximately \SI{10}{[\percent]} at $\gls{number}_{\gls{cluster}} = 50$ with  the agglomerative clustering algorithm.
With turbulent diffusivity suppressed, the turbulent contribution to species transport is negligible and there are no turbulently driven mixing zones to be bisected by compartment boundaries. Thus, the second degradation mechanism (discussed in the previous paragraphs) is entirely absent.
This confirms that the error increase in the baseline case (with turbulent mixing) was driven by a suppression of (turbulent) mixing rates in our compartment model.

A second observation further supports this interpretation:
in the baseline case, including the turbulent kinetic energy $\overline{\gls{tke}}^{\gls{liq}}$ as a clustering feature stabilised the error by guiding compartment boundaries away from regions of high turbulent intensity.
With turbulent diffusivity suppressed, using $\overline{\gls{tke}}^{\gls{liq}}$ as a feature no longer provides any additional benefit. The multi-feature combinations and the single dissolved oxygen concentration feature converge to a similar accuracy, as the turbulent structure of the flow no longer influences the concentration distribution.
As can be seen in \autoref{fig:t1875_error}, our results for all feature sets remain below the single-compartment baseline of \SI{22}{[\percent]}. This indicates that at least a moderate increase in predictive powder can be obtained with even the simplest feature (i.e., the local gas volume fraction) if turbulent mixing is non-existing.

\begin{figure}[H]
        \centering
        \begin{subfigure}[h]{0.30\columnwidth}
            \centering
            \includegraphics[width=\linewidth]{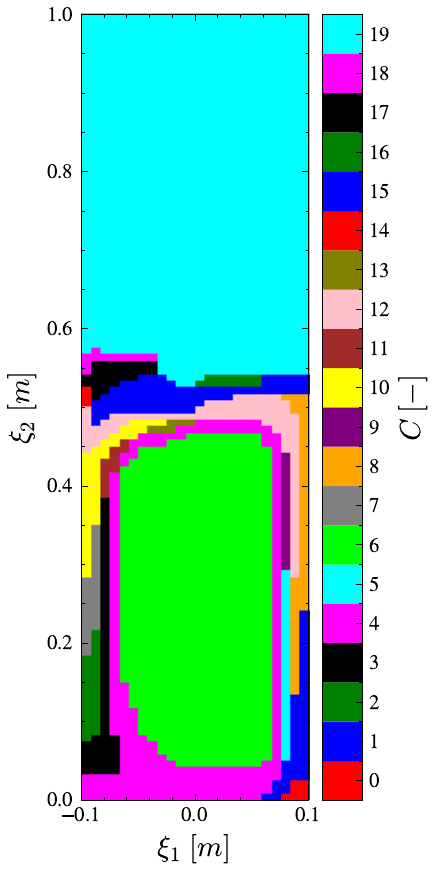}
            \caption{Agglomerative clustering}
            \label{fig:t1875_cluster_agglo}
        \end{subfigure}%
        \begin{subfigure}[h]{0.30\columnwidth}
            \centering
            \includegraphics[width=\linewidth]{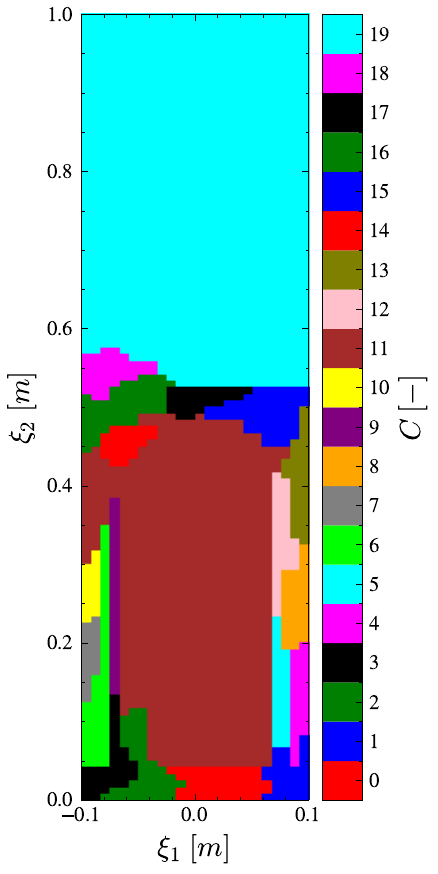}
            \caption{K-means clustering}
            \label{fig:t1875_cluster_kmeans}
        \end{subfigure}%

        \caption[Cluster geometry at $\gls{number}_{\gls{cluster}} = 20$ for the suppressed-diffusion case]{Cluster assignment at $\gls{number}_{\gls{cluster}} = 20$ using $\overline{\gls{conc}}^{\gls{liq}}_{O_2}$ as the clustering feature. Agglomerative clustering produces spatially coherent, concentric shells around the depletion zone. K-means with graph reassignment yields fragmented, geometrically irregular compartments that lack physical interpretability.}
        \label{fig:t1875_clusters}
    \end{figure}

The shape of the compartments at $\gls{number}_{\gls{cluster}} = 20$ for both algorithms are compared in \autoref{fig:t1875_clusters}, providing a direct visual explanation of their different error behaviours.

The result for the agglomerative clustering algorithm shown in panel (\ref{fig:t1875_cluster_agglo}) clearly exhibits the thin structures described above: the compartments form concentric, onion-like bands around the central depleted core, each following an oxygen iso-concentration surface.
While this geometry is expected given the use of a concentration-only feature, the shells are spatially coherent and contiguous, which preserves physical interpretability and allows the compartment model to compute meaningful inter-compartmental flow rate.

In contrast, the k-means clustering result shown in \autoref{fig:t1875_cluster_kmeans} reveals a distinct failure mode of this algorithm when combined with graph reassignment: a single compartment dominates the depleted core as a large, irregular block. Furthermore, the surrounding clusters exhibit an isotropic, spherical geometry, which is suboptimal for representing regions that are not well-mixed. Consequently, agglomerative clustering is the preferable algorithm for poorly-mixed reactors with complex concentration fields, as it inherently constructs spatially contiguous compartments by design. K-means, however, remains advantageous for highly well-mixed systems, where the concentration field is diffuse and spherical compartments appropriately reflect the underlying physics.

Finally, a third contribution to the overall error arises from the non-linear kinetics. Because the reaction rate depends non-linearly on concentration, spatial averaging over a compartment introduces an inherent mathematical discrepancy: $\langle \gls{conc}^{0.5} \rangle \neq \langle \gls{conc} \rangle^{0.5}$ \cite{fox_computational_2003}.
This closure error is unavoidable in any lumped-parameter model and becomes more significant as the intra-compartment concentration variance increases.
Despite these combined sources of error, \gls{clara} predicts the total reaction rate with an accuracy of less than \SI{2}{[\percent]}, as shown in \autoref{fig:t1871_error_R}. This high accuracy is primarily because the system is mass-transfer limited; consequently, the overall reaction rate is governed by the interphase mass transfer rate. This demonstrates that the integral quantity of engineering relevance remains well-captured, even when the local concentration field is imperfectly resolved. Extended error metrics are provided in \ref{app:extended_results} for the interested reader.

\section{Conclusion and Outlook} \label{sec:conclustion_outlook}

\subsection{Conclusion}

Our present study introduced \gls{clara}, an open-source toolbox that generates compartment models from multiphase \gls{cfd} data through unsupervised clustering. Two algorithms are implemented: k-means with a graph reassignment step and hierarchical agglomerative clustering with connectivity constraints. A robust volume flow optimisation enforces strict mass conservation within the resulting compartment network. The framework was verified against analytical solutions, validated against reactive two-phase \gls{of} simulations of a bubble column reactor, and systematically evaluated across three reaction orders, six clustering features, and up to 50 compartments.

Analytical verification against constant, linear, quadratic, and two-phase mass transfer (with constant) flow profiles confirmed that the compartment model error decreases with increasing compartment count. This result establishes the mathematical correctness of our framework.

The quarter-gassed case for the bubble column reactor, in which gas injection is confined to one side of the bottom region, generates a strong spatial heterogeneity and constitutes the primary benchmark challenge. Among the three reaction orders, the 0.5\textsuperscript{th}-order case is the most demanding: the resulting concentration field spans over two orders of magnitude and is highly sensitive to the clustering strategy, whereas first- and second-order profiles remain nearly homogeneous and yield low errors regardless of the approach. Feature selection proved to be the single most influential design decision. Purely flow-based features, such as liquid volume fraction and turbulent kinetic energy, are insensitive to concentration gradients and cannot guide the algorithm to a meaningful compartment structure. The dissolved oxygen concentration feature achieves the lowest error in this study, $\gls{error}_{\gls{conc}} \approx \SI{4.5}{[\percent]}$ at $\gls{number}_{\gls{cluster}} = 5$ with agglomerative clustering, well below the conventional single-compartment baseline of \SI{13.5}{[\percent]}. Multi-feature combinations that include both concentration and turbulent kinetic energy provide the most robust performance across a wide range of compartment numbers.

A counter-intuitive increase in local concentration error with increasing compartment number was identified and attributed to two compounding mechanisms: First, concentration-based features cause the algorithm to generate thin iso-concentration compartments at high $\gls{number}_{\gls{cluster}}$. These cluster span physically distinct zones, the gassed plume and the gas-free return flow, and misplace interfacial mass transfer into regions that are almost void of dissolved oxygen. Second, increasing the compartment number causes boundaries to bisect zones that are mixed by turbulent motion. The resulting truncation of turbulent transport, which is not captured by the net convective flows considered by or compartment model, introduces artificial concentration gradients. A numerical experiment with $\gls{schmidt_number_turbulent} = 10^7$ (which suppresses turbulent diffusivity in our CFD reference data) shows that the error decreases for such a situation with increasing cluster count. Consequently, a practical sweet spot between three and ten compartments is identified, where the well-mixed assumption remains valid and spatial resolution is not eroded by the non-existence of turbulent mixing in our compartment model.

\gls{clara} accurately captures the local concentration field while predicting the total reactor reaction rate to within \SI{2}{[\percent]}. This demonstrates that the integral quantities of primary engineering relevance are well-preserved. These results establish \gls{clara} as a robust framework for the automated generation of physically meaningful compartment models from multiphase \gls{cfd} data. This offers computationally efficient surrogates for numerous engineering applications, including (i) the efficient simulation of processes with disparate time scales, (ii) digital twin development, and (iii) real-time control of complex multiphase flow processes.

\subsection{Outlook}

The most immediate priority for future work is improved feature engineering. Residence time, as employed in combustion applications \cite{savarese_machine_2023}, encode both transport and reaction history in a single scalar and represent a physically motivated alternative to concentration-based features. However, computing residence times in a gas-fed semi-batch reactor requires dedicated tracer simulations, and extending this concept to the dispersed gas phase is non-trivial. Physics-informed composite features combining concentration, turbulent kinetic energy, and local Hatta number could provide more balanced partitioning and avoid badly placed cluster. Additionally, applying feature weighting, such as through reinforcement learning, could further improve the clustering results.

A second priority is the accurate modelling of turbulent transport rates across compartment boundaries. The high-$\gls{schmidt_number_turbulent}$ simulation experiments confirmed that turbulent species transport is the primary driver of the error increase for high compartment numbers. Although flow-informed approaches have been proposed for single-phase or structured compartments to address this turbulent exchange deficit \cite{delafosse_cfd-based_2014, maldonado_de_leon_dynamic_2025, le_nepvou_de_carfort_flow-informed_2026}, extending these methods to arbitrarily shaped two-phase compartments requires further development and remains a primary direction for future work.

Validation on a three-dimensional bioreactor geometry is the next logical step toward industrial application. Three-dimensional flows exhibit more complex recirculation patterns and turbulence structures compared to the two-dimensional bubble column studied in our present work. Investigating how our practical guidelines for feature selection and clustering algorithm choice adapt to these advanced geometries represents a highly promising avenue for future research. 


\section{Nomenclature}

\setglossarystyle{alttree}

\newcommand{\printUnitPostDesc}{%
    \ifglshasfield{unit}{\glscurrententrylabel}%
    {\hfill \makebox[5em][r]{\glsentryunit{\glscurrententrylabel}}}%
    {}%
}
\renewcommand*{\glsxtrpostdescription}{\printUnitPostDesc}

{ 
    \glsfindwidesttoplevelname[abbreviations]
    
    \renewcommand*{\glsxtrpostdescription}{}
    
    \printunsrtglossary[type=abbreviations, title={Abbreviations}]
} 

{
    \glsfindwidesttoplevelname[latinsymbols]
    
    \printunsrtglossary[type=latinsymbols, title={Latin Characters}]
}

{
    \glsfindwidesttoplevelname[greeksymbols]
    \printunsrtglossary[type=greeksymbols, title={Greek Characters}]
}

{
    \glsfindwidesttoplevelname[operatorsymbols]
    \renewcommand*{\glsxtrpostdescription}{}
    \printunsrtglossary[type=operatorsymbols, title={Operators}]
}

\section*{Acknowledgments}
Funding: This work was supported by Graz University of Technology through the "DigiBioTech" project (Lead: Prof. Robert Kourist)\\
We also thank Daniel Berger for his valuable work on the Reinforcement Learning strategy for the compartmentalization, especially on the feature selection, as well as Christoph Griesbacher for his valuable insight about clustering algorithms and ML strategies.

\section*{Declaration of competing interest}

The authors declare that they have no known competing financial interests or personal relationships that could have appeared to influence the work reported in this paper.

\section*{Declaration of generative AI and AI-assisted technologies in the manuscript preparation process}

During the preparation of this work the authors used Claude (Anthropic) and Gemini (Google) in order to assist with language editing, improving readability, and drafting portions of the manuscript text. After using this tool, the authors reviewed and edited the content as needed and take full responsibility for the content of the published article.

\section*{CRediT authorship contribution statement}
\textbf{Michael Mitterlindner:} Conceptualization, Methodology, Software, 
Formal analysis, Investigation, Validation, Visualization, Writing - original draft.
\textbf{Maximilian Graber:} Methodology, Software, Writing - review and editing.
\textbf{Regina Kratzer:} Project administration, Supervision, Writing - review and editing.
\textbf{Markus Reichhartinger:}  Project administration, Supervision, Writing - review and editing.
\textbf{Stefan Radl:} Conceptualization, Funding acquisition, Project administration, 
Supervision, Writing - review and editing.

\bibliographystyle{elsarticle-num} 
\bibliography{references}







\appendix

\section{Simulation Parameters}\label{app:simulation_parameters}
\autoref{tab:simulation_parameters} shows the simulation parameters. These parameters were also used for the compartment model. We also assume same heat capacity of oxigen in the liquid and in the gas phase so that temperature canges can be neglected.

\begin{table}[H]
    \centering
    \caption[Simulation parameters]{Simulation parameters}
    \label{tab:simulation_parameters}
    \begin{tabular}{cl}
        \toprule 
        Variable       & Value\\
        \midrule 
            $\gls{velo}_{\gls{in}}$ & \SI{0.05}{[\meter\per\second]} \\
            $\gls{lewis_number}$ & \SI{70}{[-]}\\ 
            $\gls{schmidt_number_turbulent}$ & \SI{0.7}{[-]} \\ 
            $\gls{prantl_number}^{\gls{liq}}$ & \SI{7.0}{[-]}\\ 
            $\gls{prantl_number}^{\gls{gas}}$ & \SI{0.7}{[-]}\\
            $\gls{mol_weight}_{O_2}$ & \SI{32.0}{[\kilo\gram\per\kilo\mole]}  \\ 
            $\gls{mol_weight}_{H_2O}$ & \SI{18.0}{[\kilo\gram\per\kilo\mole]}  \\
            $\gls{dynamic_visc}^{\gls{liq}}$ & \SI{1e-3}{[\pascal \ \second]} \\ 
            $\gls{diameter}^{\gls{liq}}$ & \SI{1e-4}{[\meter]}  \\ 
            $\gls{diameter}^{\gls{gas}}$ & \SI{3e-3}{[\meter]}  \\
            $\gls{pressure}_{\gls{tot}}$ & \SI{1e5}{[\pascal]}  \\ 
            $\gls{temp}$ & \SI{293.15}{[\kelvin]}  \\ 
            $\gls{henry_const}^*_{O_2 \ in \ H_2O}$ & \SI{3.45e-2}{[-]} ,\cite{sander_compilation_2023}    \\ 
            $\gls{density}^{\gls{liq}}$ & \SI{1e3}{[\kilo\gram\per\cubic\meter} \\ 
            $\gls{density}^{\gls{gas}}$ (calculated) & \SI{1.31}{[\kilo\gram\per\cubic\meter]} \\ 
            resolution & \SI{120}{[nodes\per\meter]} \\
            domain size & 0.2 x 1 x 0.1\si{[\meter]} \\
            turbulence & $k\epsilon$-model \\
            solver & reactingTwoPhaseEulerFoam \\
            drag model & Schiller-Naumann, \cite{schiller_uber_1933}\\
            mass transfer model & Frössling, \cite{frossling_uber_1938}\\
            
        \bottomrule 
    \end{tabular}

\end{table}

\section{Analytical Solutions}\label{app:analytical_solutions}

\begin{figure}[!htb]
    \centering
    \includegraphics[width=\columnwidth]{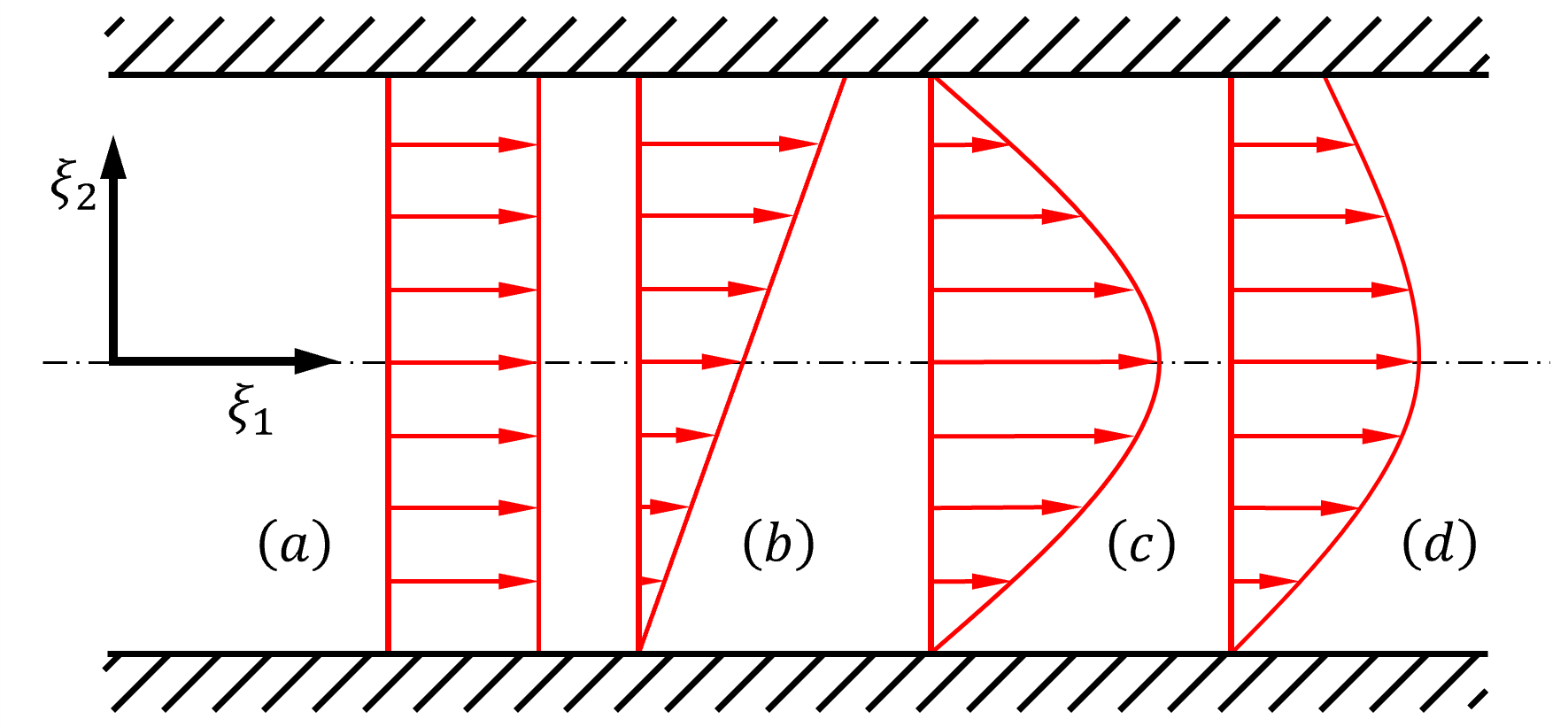}
    \caption{Schematic representation of the flow profiles considered for analytical verification: (a) plug, (b) linear,  (c) quadratic, and  (d) superposition of the linear and quadratic flow profile.}
    \label{fig:flow_profiles_sandbox}
\end{figure}

In this section, analytical solutions are derived for the concentration distribution of a chemical species undergoing a first-order reaction within a two-dimensional channel flow between two infinite parallel plates. These analytical benchmarks are essential for the verification of the numerical models presented later in this work. Four distinct flow profiles are considered: plug, linear (Couette), quadratic (Poiseuille), and a superposition of the linear and quadratic flow profile, as illustrated in \autoref{fig:flow_profiles_sandbox}.

\subsection{Single-Phase Analytical Theory}

    We consider a simple irreversible first-order chemical reaction of the form:
    \begin{equation}
        \gls{stoich_coeff}_0 \gls{conc}_0 \xrightarrow{\gls{rrc1O}} \gls{stoich_coeff}_1 \gls{conc}_1
        \label{eq:chemical_reaction}
    \end{equation}
    
    where $\gls{conc}_0$ is the reactant concentration and $\gls{rrc1O}$ is the reaction rate constant. The steady-state transport of the reactant, governed by advection and reaction, is described by:
    
    \begin{equation}
        \gls{velo_vec} \cdot \nabla \gls{conc} = -\gls{rrc1O} \gls{conc}
        \label{eq:advection_reaction}
    \end{equation}
    
    where $\gls{velo_vec}$ is the velocity vector. For the analytical derivation, we assume a high Péclet number regime where diffusive transport is negligible compared to advection.
    
    To generalize the results, we introduce the following dimensionless variables:
    $
        \gls{conc}^* = \frac{\gls{conc}}{\gls{conc}_{\gls{in}}}, \quad \gls{coord}_1^* = \frac{\gls{coord}_1}{\gls{length_channel}}, \quad \gls{coord}_2^* = \frac{\gls{coord}_2}{\gls{height}}, \quad \gls{velo_vec}^* = \frac{\gls{velo_vec}}{\langle \gls{velo}_1 \rangle_{\gls{coord}_2}}
    $
    where $\gls{conc}_{\gls{in}}$ is the inlet concentration, $\gls{length_channel}$ is the channel length, $\gls{height}$ is the channel height, and $\langle \gls{velo}_1 \rangle_{\gls{coord}_2}$ is the mean velocity. The velocity profile is assumed to be fully developed and unidirectional, i.e., $\gls{velo_vec}^* = [\gls{velo}_1^*(\gls{coord}_2^*), 0]^T$. We note in passing that we solve the equation for the educt concentration $\gls{conc}_0$ and therefore the inlet concentration is not zero. The product concentration can be claculated as follows: $\gls{conc}_{1} = \gls{conc}_{\gls{in}, 1} + \frac{\gls{stoich_coeff}_0}{\gls{stoich_coeff}_1}\gls{conc}_0$ 
    
    Substituting these into \autoref{eq:advection_reaction}, we obtain the dimensionless governing equation:
    \begin{equation}
        \gls{velo}_1^*(\gls{coord}_2^*) \frac{\partial \gls{conc}^*}{\partial \gls{coord}_1^*} = -\gls{Da} \, \gls{conc}^*
        \label{eq:advection_reaction_dimless}
    \end{equation}
    where $\gls{Da} = \frac{\gls{rrc1O} \gls{length_channel}}{\langle \gls{velo}_1 \rangle_{\gls{coord}_2}}$ is the Damköhler number, representing the ratio of the residence time to the reaction time scale. Integrating \autoref{eq:advection_reaction_dimless} with respect to $\gls{coord}_1^*$ subject to the boundary condition $\gls{conc}^*(0, \gls{coord}_2^*) = 1$ yields the general solution:
    \begin{equation}
        \gls{conc}^*(\gls{coord}_1^*, \gls{coord}_2^*) = \exp\left( -\frac{\gls{Da} \cdot \gls{coord}_1^*}{\gls{velo}_1^*(\gls{coord}_2^*)} \right)
        \label{eq:general_solution}
    \end{equation}

\subsubsection{Analytical Solutions for Specific Flow Regimes}

    \paragraph{Constant (Plug) Flow Profile}

    For ideal plug flow, the velocity is uniform across the channel height. The dimensionless velocity profile and mean velocity are given by:
    $\gls{velo}_1^*(\gls{coord}_2^*) = 1$, $\langle \gls{velo}_1 \rangle_{\gls{coord}_2} = \gls{velo}_{\gls{max}}$
    The Damköhler number, for this flow profile, is defined as $\gls{Da} = \frac{\gls{rrc1O} \gls{length_channel}}{\gls{velo}_{\gls{max}}}$. Substituting the velocity profile into \autoref{eq:general_solution}, the concentration field is:
    
    \begin{equation}
        \gls{conc}^*(\gls{coord}_1^*) = \exp \left(  {-\gls{Da} \gls{coord}_1^*} \right)
        \label{eq:plug_flow_solution}
    \end{equation}

    \paragraph{Linear (Couette) Flow Profile}
    
    The linear flow profile comes from moving  the top plate relative to the stationary bottom plate. The dimensionless velocity profile is linear:
    \begin{equation}
        \gls{velo}_1^*(\gls{coord}_2^*) = \gls{coord}_2^* + \frac{1}{2}
        \label{eq:linear_velocity_profile}
    \end{equation}
    assuming the coordinate system origin is at the channel center. The mean velocity is $\langle \gls{velo}_1 \rangle_{\gls{coord}_2} = \gls{velo}_{\gls{max}}/2$, leading to a Damköhler number of $\gls{Da} = \frac{2 \gls{rrc1O} \gls{length_channel}}{\gls{velo}_{\gls{max}}}$. The resulting concentration distribution is:
    \begin{equation}
        \gls{conc}^*(\gls{coord}_1^*, \gls{coord}_2^*) = \exp\left( - \frac{\gls{Da} \cdot \gls{coord}_1^*}{\gls{coord}_2^* + 1/2} \right)
        \label{eq:couette_flow_solution}
    \end{equation}

    \paragraph{Quadratic (Poiseuille) Flow Profile}
    
    For pressure-driven flow between parallel plates, the velocity profile is quadratic (parabolic):
    \begin{equation}
        \gls{velo}_1^*(\gls{coord}_2^*) = 1 - 4{\gls{coord}_2^*}^2
        \label{eq:quadratic_velocity_profile}
    \end{equation}
    The mean velocity is $\langle \gls{velo}_1 \rangle_{\gls{coord}_2} = \frac{2}{3}\gls{velo}_{\gls{max}}$, resulting in $\gls{Da} = \frac{3 \gls{rrc1O} \gls{length_channel}}{2\gls{velo}_{\gls{max}}}$. The concentration field is given by:
    \begin{equation}
        \gls{conc}^*(\gls{coord}_1^*, \gls{coord}_2^*) = \exp\left( - \frac{\gls{Da} \cdot \gls{coord}_1^*}{1 - 4{\gls{coord}_2^*}^2} \right)
        \label{eq:poiseuille_flow_solution}
    \end{equation}

    \paragraph{Superposition of a linear and a quadratic flow profile}
    
    A generalized profile combining shear-driven and pressure-driven flow can be described by a blending factor $\gls{lqfp_blend} \in [0, 1]$. The velocity profile is:
    \begin{equation}
        \gls{velo}_1^*(\gls{coord}_2^*) = (1-\gls{lqfp_blend})(1-4{\gls{coord}_2^*}^2) + \gls{lqfp_blend}\left(\gls{coord}_2^* + \frac{1}{2}\right)
        \label{eq:superposition_velocity_profile}
    \end{equation}
    The mean velocity for this combined flow is:
    \begin{equation}
        \langle \gls{velo}_1 \rangle_{\gls{coord}_2} = \gls{velo}_{\gls{max}} \left( \frac{3 + \gls{lqfp_blend}}{6} \right)
        \label{eq:mean_velocity_superposition}
    \end{equation}
    The corresponding Damköhler number is $\gls{Da} = \frac{6 \gls{rrc1O} \gls{length_channel}}{\gls{velo}_{\gls{max}}(3+\gls{lqfp_blend})}$. The concentration field is obtained by substituting these expressions into \autoref{eq:general_solution}. We note in passing that the used $\gls{velo}_{\gls{max}}$ is not the real maximal velocity of the flow profile, it is a weighted combination of the linear max velocity and the quadratic max velocity. This was used to simplify the calculation.

\subsection{Two-Phase Analytical Theory}
Dissolution of species will be described.
\subsubsection{Concentration based Molar Transfer Rate}
    Molar flow rate (concentration based):
    \begin{equation}
        \gls{mol_flow}_i = \gls{kLa}_{i} \cdot \gls{Vol}_{\gls{tot}} \cdot (\gls{conc}^{\gls{interface}}_{i}-\gls{conc}_{i}^{\gls{liq}})
        \label{eq:interface_mole_flow}
    \end{equation}
    
    with $\gls{conc}^{\gls{interface}}_{i}$ calculated from Henry's law (again in dimensionless form) with the partial pressure ($\gls{pressure}_{i} = \gls{mol_frac}_{i} \cdot \gls{pressure}_{\gls{tot}}$)

    \begin{equation}
        \gls{conc}^{\gls{interface}}_{i} =  \frac{\gls{henry_const}^*_{i} \gls{pressure}_{i}}{\gls{gas_const} \  \gls{temp}}
        \label{eq:interface_concentration}
    \end{equation}

\subsubsection{Calculation of the maximal Solubility}

    Solubility in \si{[\kilo\mole\per\cubic\meter]}:
    \begin{equation}
        \gls{conc}^{\gls{equil}}_{i} =  \frac{\gls{henry_const}^*_{i} \gls{pressure}_{i}}{\gls{gas_const} \  \gls{temp}}
        \label{eq:equillibrium_concentration_kmolpercubicmeter}
    \end{equation}
    
    Dimensionless solubility \si{[\si{\kilo\gram} / \si{\kilo\gram}_{\gls{tot}}]}:
    \begin{equation}
        \gls{weight_frac}^{\gls{equil}}_{i} =  \frac{\gls{henry_const}^*_{i} \gls{pressure}_{i}}{\gls{gas_const} \ \gls{temp}}\cdot\frac{\gls{mol_weight}_{i}}{\gls{density}_{i}}
        \label{eq:equillibrium_concentration_kgpercubicmeter}
    \end{equation}
    
    as an example for $O_2$ in $H_2O$ with \SI[scientific-notation=true, round-mode=none]{1.41e-3}{[\kilo\mole\per\cubic\meter]} and for a mass fraction (used in Openfoam): \SI[scientific-notation=true, round-mode=none]{4.5e-5}{[\si{\kilo\gram} / \si{\kilo\gram}_{\gls{tot}}]} at \SI{293.15}{[\si{\kelvin}]} and \SI[scientific-notation=true]{1e5}{[\si{\pascal}]}.

\subsubsection{Calculation of the Volumetric Mass Transfer Coefficient}

    The determination of the volumetric mass transfer coefficient $\gls{kLa_liq}$ is based on the dynamic gassing-in method. The change in dissolved oxygen concentration $\gls{conc}^{\gls{liq}}_{O_2}$ over time is described by the mass balance equation.
    
    To convert the mass fraction $\gls{weight_frac}_{O_2}$ (output by OpenFOAM) to molar concentration $\gls{conc}^{\gls{liq}}_{O_2}$ in \si{[\kilo\mole\per\cubic\meter]} for the calculation, the following relationship is applied:

    \begin{equation}
        \gls{conc}^{\gls{liq}}_{O_2} = \gls{weight_frac}_{O_2} \cdot \frac{\gls{density}^{\gls{liq}}}{\gls{mol_weight}_{O_2}}
        \label{eq:conversion_massfrac_to_molar}
    \end{equation}
    
    Note that the concentration $\gls{conc}^{\gls{liq}}_{O_2}$ is defined per unit volume of liquid phase, not total reactor volume and therefore this $\gls{kLa_liq}$ value is also per liquid volume. For this "experimental" determination we considered just solution of the species without a liquid reaction.
    
    \begin{equation}
        \frac{d\gls{conc}^{\gls{liq}}_{O_2}}{d \gls{time}} = \gls{kLa_liq} (\gls{conc}^{\gls{equil}}_{O_2} - \gls{conc}^{\gls{liq}}_{O_2})
        \label{eq:mass_transfer_balance}
    \end{equation}
    
    where $\gls{conc}^{\gls{equil}}$ is the saturation concentration at equilibrium (calculated in Eq. \ref{eq:equillibrium_concentration_kmolpercubicmeter}). Assuming the liquid phase is initially free of oxygen ($\gls{conc}^{\gls{liq}}_{O_2}(\gls{time}=0) = 0$), integrating Equation \ref{eq:mass_transfer_balance} yields the time-dependent concentration profile:
    
    \begin{equation}
        \gls{conc}^{\gls{liq}}_{O_2}(\gls{time}) = \gls{conc}^{\gls{equil}}_{O_2} \cdot \left( 1 - e^{-\gls{kLa_liq} \cdot \gls{time}} \right)
        \label{eq:concentration_over_time}
    \end{equation}
    
    The $\gls{kLa_liq}$ value is determined by performing a non-linear least squares regression of Equation \ref{eq:concentration_over_time} to the probe data obtained from the simulation.

\subsubsection{Solution for Plug Flow with Interface Mass Transfer}

    \cite{dhaouadi_gasliquid_2008}  presents an analytical solution for gas–liquid mass transfer in a bubble column reactor, focusing on the transient response during gassing-in experiments. Assuming plug flow for the liquid phase and a steady gas phase, the model incorporates interface mass transfer governed by Henry’s law, as well as static pressure effects on gas solubility and the dynamic response of DO sensors. The authors derive a Laplace-domain solution for dissolved oxygen concentration profiles and validate it experimentally using a 6-meter-high bubble column. The model enables accurate estimation of the volumetric mass transfer coefficient \gls{kLa} and shows good agreement with numerical simulations and empirical correlations.  In our case we just derive the steady state solution.

\paragraph{Assumptions}

    The following assumptions are made for the derivation of the analytical solution:
    \begin{itemize}
        \item the system is at steady state 
        \item one-dimensional plug flow in the $\gls{coord}_1$-direction (i.e., infinite fast lateral transport)
        \item no diffusion or disperion in axial direction
        \item both phases (gas and liquid) flow co-currently with constant superficial velocities $\gls{velo}_{\gls{superficial}}^{\gls{gas}}=\gls{vol_frac}^{\gls{gas}}\gls{velo}^{\gls{gas}}$ and $\gls{velo}_{\gls{superficial}}^{\gls{liq}}=\gls{vol_frac}^{\gls{liq}}\gls{velo}^{\gls{liq}}$
        \item the gas and liquid occupy a spatially uniform volume fraction $\gls{vol_frac}^{\gls{gas}}$ and $\gls{vol_frac}^{\gls{liq}} = 1 - \gls{vol_frac}^{\gls{gas}}$ 
        \item the volumetric mass transfer coefficient \gls{kLa} is spatially uniform (e.g., assuming constant bubble size, interface area and slip velocity)
        \item species $i=0$ is a non-transferring carrier phase with constant concentration in both phases
        \item species $i=1$ is a transferring species and undergoes mass transfer governed by Henry's law
    \end{itemize}

\paragraph{Governing Equations}

The interface molar transfer rate per unit total volume is defined as:
\begin{equation}
    \gls{mol_trans_rate_per_vol}^{\gls{interface}}_1 = \gls{kLa} \left( \gls{henry_const}^*_{1} \gls{conc}_{1}^{\gls{gas}} - \gls{conc}_{1}^{\gls{liq}} \right)
    \label{eq:interphase_transfer_rate}
\end{equation}

The plug flow mass balances for species 1 are:
\begin{align}
    \gls{vol_frac}^{\gls{gas}} \gls{velo}^{\gls{gas}} \frac{d \gls{conc}_{1}^{\gls{gas}}}{d \gls{coord}} &= -\gls{mol_trans_rate_per_vol}^{\gls{interface}}_1 = -\gls{kLa} (\gls{henry_const}^*_{1} \gls{conc}_{1}^{\gls{gas}} - \gls{conc}_{1}^{\gls{liq}}) \label{eq:gas_mass_balance} \\
    \gls{vol_frac}^{\gls{liq}} \gls{velo}^{\gls{liq}} \frac{d \gls{conc}_{1}^{\gls{liq}}}{d \gls{coord}} &= +\gls{mol_trans_rate_per_vol}^{\gls{interface}}_1 = +\gls{kLa} (\gls{henry_const}^*_{1} \gls{conc}_{1}^{\gls{gas}} - \gls{conc}_{1}^{\gls{liq}}) \label{eq:liquid_mass_balance}
\end{align}
Adding the two results in :

\begin{align}
    \gls{vol_frac}^{\gls{gas}} \gls{velo}^{\gls{gas}} \frac{d \gls{conc}_{1}^{\gls{gas}}}{d \gls{coord}} + \gls{vol_frac}^{\gls{liq}} \gls{velo}^{\gls{liq}} \frac{d \gls{conc}_{1}^{\gls{liq}}}{d \gls{coord}}=  0 
    \label{eq:added_mass_balance}
\end{align}

We define the total molar flux $\gls{mol_flux}_{\gls{tot},1}$ of species 1 (in the gas AND the liquid phase there is just a redistribution happening), which is conserved along $x$ with being the integral of  \eqref{eq:added_mass_balance}:
\begin{equation}
    \gls{mol_flux}_{\gls{tot},1} = \gls{vol_frac}^{\gls{gas}} \gls{velo}^{\gls{gas}} \gls{conc}_{1}^{\gls{gas}} + \gls{vol_frac}^{\gls{liq}} \gls{velo}^{\gls{liq}} \gls{conc}_{1}^{\gls{liq}} = \gls{vol_frac}^{\gls{gas}} \gls{velo}^{\gls{gas}} \gls{conc}_{1}^{\gls{gas},0} + \gls{vol_frac}^{\gls{liq}} \gls{velo}^{\gls{liq}} \gls{conc}_{1}^{\gls{liq},0}
    \label{eq:total_flux_species1}
\end{equation}

From this, the liquid concentration can be expressed as a function of the gas concentration:
\begin{equation}
    \gls{conc}_{1}^{\gls{liq}}(\gls{coord}) = \frac{1}{\gls{vol_frac}^{\gls{liq}} \gls{velo}^{\gls{liq}}} \left( \gls{mol_flux}_{\gls{tot},1} - \gls{vol_frac}^{\gls{gas}} \gls{velo}^{\gls{gas}} \gls{conc}_{1}^{\gls{gas}}(\gls{coord}) \right)
    \label{eq:cliq_from_cgas}
\end{equation}

\paragraph{Derivation of the Solution}

Substituting \eqref{eq:cliq_from_cgas} into the gas-phase balance \eqref{eq:gas_mass_balance} yields:
\begin{align}
    \gls{vol_frac}^{\gls{gas}} \gls{velo}^{\gls{gas}} \frac{d \gls{conc}_{1}^{\gls{gas}}}{d\gls{coord}} &= -\gls{kLa} \left( \gls{henry_const}^*_{1} \gls{conc}_{1}^{\gls{gas}} - \frac{1}{\gls{vol_frac}^{\gls{liq}} \gls{velo}^{\gls{liq}}} (\gls{mol_flux}_{\gls{tot},1} - \gls{vol_frac}^{\gls{gas}} \gls{velo}^{\gls{gas}} \gls{conc}_{1}^{\gls{gas}}) \right) \label{eq:gas_balance_substituted} \\
    &= -\gls{kLa} \left[ \left( \gls{henry_const}^*_{1} + \frac{\gls{vol_frac}^{\gls{gas}} \gls{velo}^{\gls{gas}}}{\gls{vol_frac}^{\gls{liq}} \gls{velo}^{\gls{liq}}} \right) \gls{conc}_{1}^{\gls{gas}} - \frac{\gls{mol_flux}_{\gls{tot},1}}{\gls{vol_frac}^{\gls{liq}} \gls{velo}^{\gls{liq}}} \right] \label{eq:gas_balance_expanded}
\end{align}

We define the following intermediate parameters:
\begin{align}
    1/\gls{length_ref} &= \frac{\gls{kLa}}{\gls{vol_frac}^{\gls{gas}} \gls{velo}^{\gls{gas}}} \left( \gls{henry_const}^*_{1} + \frac{\gls{vol_frac}^{\gls{gas}} \gls{velo}^{\gls{gas}}}{\gls{vol_frac}^{\gls{liq}} \gls{velo}^{\gls{liq}}} \right) \label{eq:def_intermediat_params} \\
    \gls{inter_para_tpmte} &= \frac{\gls{kLa} \gls{mol_flux}_{\gls{tot},1}}{\gls{vol_frac}^{\gls{gas}} \gls{velo}^{\gls{gas}} \gls{vol_frac}^{\gls{liq}} \gls{velo}^{\gls{liq}}} \label{eq:b_def}
\end{align}

Thus, the important length scale is:
\begin{align}
    \gls{length_ref} &= \frac{\gls{vol_frac}^{\gls{gas}} \gls{velo}^{\gls{gas}}}
                   {\gls{kLa} \left( \gls{henry_const}^*_{1} + \frac{\gls{vol_frac}^{\gls{gas}} \gls{velo}^{\gls{gas}}}{\gls{vol_frac}^{\gls{liq}} \gls{velo}^{\gls{liq}}} \right) }
    \label{eq:def_length_ref}
\end{align}

In most practical applications, one uses volumetric flow rates instead of phase velocities, and therefore also the dimensionless volumetric flow of gas per volumetric flow of liquid can be defined as $\gls{vol_flow}^*$:
\begin{equation}
     \frac{\gls{vol_frac}^{\gls{gas}}\gls{velo}^{\gls{gas}}}
          {\gls{vol_frac}^{\gls{liq}}\gls{velo}^{\gls{liq}}}
    =\frac{ \gls{vol_flow}^{\gls{gas}} }
          { \gls{vol_flow}^{\gls{liq}}}    = \gls{vol_flow}^*
    \label{eq:ratioVolumetricFlowRates}
\end{equation}

The equation becomes a linear \gls{ode}:
\begin{equation}
    \frac{d \gls{conc}_{1}^{\gls{gas}}(\gls{coord})}{d \gls{coord}} + \frac{\gls{conc}_{1}^{\gls{gas}}(\gls{coord})}{\gls{length_ref}}  = \gls{inter_para_tpmte}
    \label{eq:linear_ode}
\end{equation}

The equation can be made dimensionless as follows:
\begin{equation}
    \frac{d \gls{conc}_{1}^{\gls{gas}}(\gls{coord})/\gls{conc}_{1}^{\gls{gas},0}}{d\gls{coord}/\gls{length_ref}} + \frac{\gls{conc}_{1}^{\gls{gas}}(\gls{coord})}{\gls{conc}_{1}^{\gls{gas},0}}  = \frac{\gls{inter_para_tpmte} \gls{length_ref}}{\gls{conc}_{1}^{\gls{gas},0}}
    \label{eq:ode_dimensionless_1}
\end{equation}

\begin{equation}
    \frac{d \gls{conc}_{1}^{\gls{gas}}(\gls{coord})/\gls{conc}_{1}^{\gls{gas},0}}{d\gls{coord}/\gls{length_ref}} + \frac{\gls{conc}_{1}^{\gls{gas}}(\gls{coord})}{\gls{conc}_{1}^{\gls{gas},0}}  = \gls{inter_para_tpmte_dimless}
    \label{eq:ode_dimensionless_2}
\end{equation}

with the new parameter $\gls{inter_para_tpmte_dimless}$ defined as:
\begin{equation}
    \gls{inter_para_tpmte_dimless} 
    = \left( \gls{vol_flow}^*
    +\frac{\gls{conc}_{1}^{\gls{liq},0}}{\gls{conc}_{1}^{\gls{gas},0}}    \right)
    \cdot 
    \frac{1}
         {\gls{henry_const}^*_{1}+\gls{vol_flow}^*}
    \label{eq:mu_definition}
\end{equation}

This equation has the solution, with $\gls{coord}^*=\gls{coord} / \gls{length_ref}$:
\begin{equation}
    \gls{conc}_{1}^{\gls{gas}}(\gls{coord})= 
    \gls{conc}_{1}^{\gls{gas},0} \left( e^{-\gls{coord}^*} + \gls{inter_para_tpmte_dimless} (1-e^{-\gls{coord}^*})  \right)
    \label{eq:cgas_solution}
\end{equation}

For the liquid phase, we can derive the concentration profile by substituting \eqref{eq:cgas_solution} into \eqref{eq:cliq_from_cgas} gives:
\begin{equation}
    \gls{conc}_{1}^{\gls{liq}}(\gls{coord}) = 
    \gls{conc}_{1}^{\gls{liq},0} +  \gls{vol_flow}^*\gls{conc}_{1}^{\gls{gas},0}
    \left[ 1-\left( e^{-\gls{coord}^*} + \gls{inter_para_tpmte_dimless} (1-e^{-\gls{coord}^*})  \right) \right]
    \label{eq:cliq_solution}
\end{equation}

The transferred mass per unit (total) reactor volume $\gls{Vol}_{\gls{tot}}$ can be calculated from the gas or liquid phase concentration profile. For the former, this yields:
\begin{equation}
    \frac{\gls{mol_flow}^{\gls{gas}}_{trans,1}}{\gls{Vol}_{\gls{tot}}}
    = \left(\gls{conc}_{1}^{\gls{gas}}(\gls{coord}) - \gls{conc}_{1}^{\gls{gas},0} \right)
     \frac{ \gls{vol_frac}^{\gls{gas}} \gls{velo}^{\gls{gas}}}{\gls{coord}}
    \label{eq:cg_solution}
\end{equation}

\begin{equation}
    \frac{\gls{mol_flow}^{\gls{liq}}_{trans,1}}{\gls{Vol}_{\gls{tot}}}
    = \left(\gls{conc}_{1}^{\gls{liq}}(\gls{coord}) - \gls{conc}_{1}^{\gls{liq},0}  \right)
     \frac{ \gls{vol_frac}^{\gls{liq}} \gls{velo}^{\gls{liq}}}{\gls{coord}}
    \label{eq:cl_solution}
\end{equation}

for $\gls{coord}$ the reactor length has to be inserted

\textbf{Limiting case 1: very high gas flow rate}

The length scale $\gls{length_ref}$ becomes:
\begin{equation}
    \gls{length_ref}
    =\frac{ \gls{vol_frac}^{\gls{liq}}\gls{velo}^{\gls{liq}} }
          { \gls{kLa} }    
    \label{eq:limitingCase1Lref}
\end{equation}

The parameter $\gls{inter_para_tpmte_dimless}$ becomes:
\begin{equation}
    \gls{inter_para_tpmte_dimless}
    =1    
    \label{eq:limitingCase1mu}
\end{equation}
Hence, the gas phase concentration remains constant and equal to $\gls{conc}_{1}^{\gls{gas},0}$
as $\gls{inter_para_tpmte_dimless}$ is one the two exponential therms zero out and the result is $\gls{conc}_{1}^{\gls{gas},0}$.

\textbf{Limiting case 2: rapid mass transfer}
The length scale $\gls{length_ref}$ becomes zero, and hence the exponential terms vanish. Hence, the result for the
dimensionless gas concentration profile is simply $\gls{inter_para_tpmte_dimless}$.



\section{Additional Error Metrics}\label{app:extended_results}
For the sake of brevity, we show only additional results for the 0.5th-order case. All other results with the respective plots can be found in the supplementary material.

\subsection{Additional Results}
As it is difficult to define a representative error for this specific case setup, we discuss three additional error metrics in this appendix.

\subsubsection{Concentration Averages and Performance Calculation}
For the concentration of the compartment model, we use the final concentration of the initial value problem:
\begin{equation}
    \langle \gls{conc}_{\gls{cm}} \rangle = \frac{\sum \gls{conc_vec}(\gls{time}_{end})_{O_2}^{\gls{liq}} \circ \gls{Vol_vec}^{\gls{liq}}}{\sum \gls{Vol_vec}^{\gls{liq}}}
    \label{eq:conc_avg_cm}
\end{equation}
The concentration from the simulation is calculated analogously, using the time-averaged concentration of the last averaging interval.
\begin{equation}
    \langle \gls{conc}_{\gls{cfd}} \rangle = \frac{\sum \overline{\gls{conc_vec}}_{O_2}^{\gls{liq}} \circ \gls{vol_frac_vec}^{\gls{liq}} \circ \gls{Vol_vec}}{\sum\gls{vol_frac_vec}^{\gls{liq}} \circ \gls{Vol_vec}}
    \label{eq:conc_avg_cfd}
\end{equation}

As a reference, a single-compartment conventional model (without CFD) is employed, governed by:
\begin{equation}
    \frac{d\gls{conc}^{\gls{liq}}}{d\gls{time}} = \gls{kLa_liq}\cdot \left( \gls{conc}^{\gls{equil}} - \gls{conc}^{\gls{liq}} \right) - \gls{rrc} \cdot \gls{conc}^{\gls{react_order}}
    \label{eq:onecomp_model}
\end{equation}

The equilibrium concentration is calculated as follows: $\gls{conc}^{\gls{equil}} = \gls{henry_const}^{*} \ \gls{conc}^{\gls{gas}}_{\gls{in}}$.
This model can then be solved with an initial value problem solver, e.g.\ Runge--Kutta, assuming a dissolved starting concentration of 0.

\subsubsection{Reactor Performance}

The total molar reaction rate is calculated as follows. For \gls{clara}:
\begin{equation}
    \gls{mol_rate}^{\gls{tot}}_{\gls{cm}} = \sum \gls{rrc} \circ \left( \gls{conc_vec}(\gls{time}_{end})_{O_2}^{\gls{liq}} \right)^{\gls{react_order}} \circ \gls{Vol_vec}^{\gls{liq}}
    \label{eq:mol_rate_cm}
\end{equation}
From \gls{cfd}:
\begin{equation}
    \gls{mol_rate}^{\gls{tot}}_{\gls{cfd}} = \sum \gls{rrc} \circ \left( \overline{\gls{conc_vec}}_{O_2}^{\gls{liq}} \right)^{\gls{react_order}} \circ \gls{vol_frac_vec}^{\gls{liq}} \circ \gls{Vol_vec}
    \label{eq:mol_rate_cfd}
\end{equation}

The relative reactor performance error is defined as:
\begin{equation}
    \gls{error}_{\gls{mol_rate}} = \frac{\gls{mol_rate}^{\gls{tot}}_{\gls{cm}} - \gls{mol_rate}^{\gls{tot}}_{\gls{cfd}}}{\gls{mol_rate}^{\gls{tot}}_{\gls{cfd}}}
    \label{eq:error_mol_rate}
\end{equation}

Another way to represent the \gls{cm} error, is to use the liquid-volume-weighted \gls{rmse}. Here, $\gls{conc}_{map}$ is the compartment concentration mapped back onto the \gls{cfd} grid, and $\gls{conc}^*_{floor}$ ($\gls{conc}_{floor} = \gls{conc}_{\gls{equil}}\gls{conc}^*_{floor}$) is a floor concentration introduced to avoid large relative errors for immeasurably low concentrations and to prevent zero-argument errors in the logarithmic metrics discussed in Appendix~\ref{app:extended_results}. The dimensionless floor concentration is set to \SI{1e-4}{[-]}, as oxygen concentrations below this threshold would already be difficult to measure experimentally. For $\gls{error}_{\gls{rmse}}$, only cells with a liquid volume fraction above 0.6 are considered, as these constitute the bulk liquid region of interest.

\begin{equation}
    \gls{error}_{\gls{rmse}} = \sqrt{
    \frac{ \sum{ \gls{Vol_vec}^{\gls{liq}} \left( \frac{\gls{conc_vec}_{\gls{cfd}} - \gls{conc_vec}_{map}} { \gls{conc_vec}_{\gls{cfd}} + \gls{conc}_{floor}} \right)^2  }}
    {\sum{\gls{Vol_vec}^{\gls{liq}}}}
    }
    \label{eq:error_rmse}
\end{equation}

\begin{equation}
    \gls{error}_{\gls{rmse}}^{log} = \sqrt{
    \frac{ \sum{ \gls{Vol_vec}^{\gls{liq}} \left( \log(\max(\gls{conc_vec}_{\gls{cfd}},  \gls{conc}_{floor})) - \log(\max(\gls{conc_vec}_{map},  \gls{conc}_{floor})) \right)^2  }}
    {\sum{\gls{Vol_vec}^{\gls{liq}}}}
    }
    \label{eq:error_rmse_log}
\end{equation}

\subsubsection{Additional Results for the 0.5th-Order Case}

\begin{figure}[H]
        \begin{subfigure}[b]{0.3\columnwidth}
            \centering
            \includegraphics[width=\linewidth]{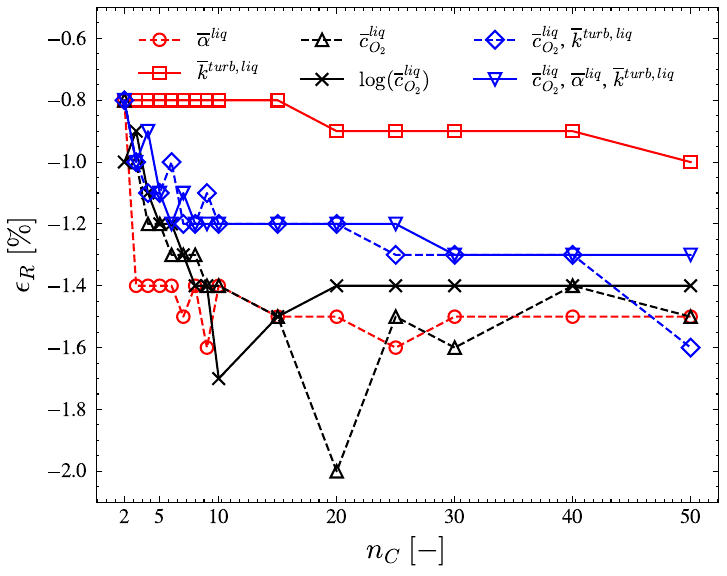}
            \caption{$\gls{error}_{\gls{mol_rate}}$}
            \label{fig:t1871_error_R}
        \end{subfigure}
        \begin{subfigure}[b]{0.3\columnwidth}
            \centering
            \includegraphics[width=\linewidth]{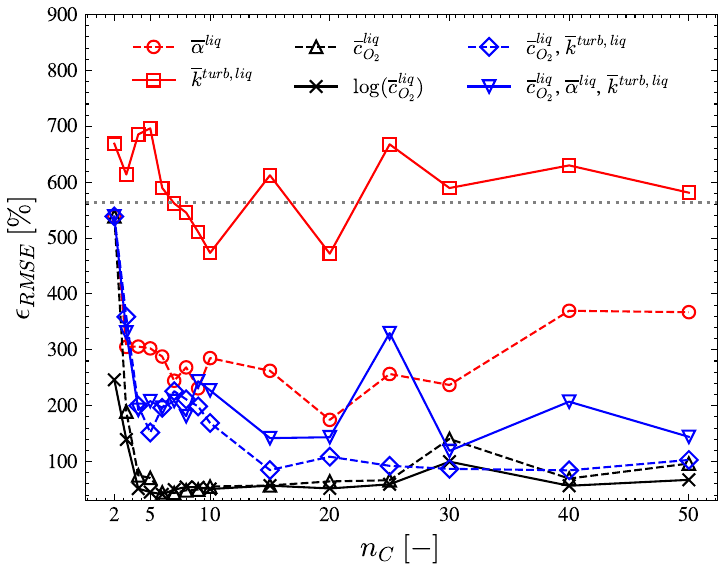}
            \caption{$\gls{error}_{\gls{rmse}}$}
            \label{fig:t1871_error_RMSE}
        \end{subfigure}
        \begin{subfigure}[b]{0.3\columnwidth}
            \centering
            \includegraphics[width=\linewidth]{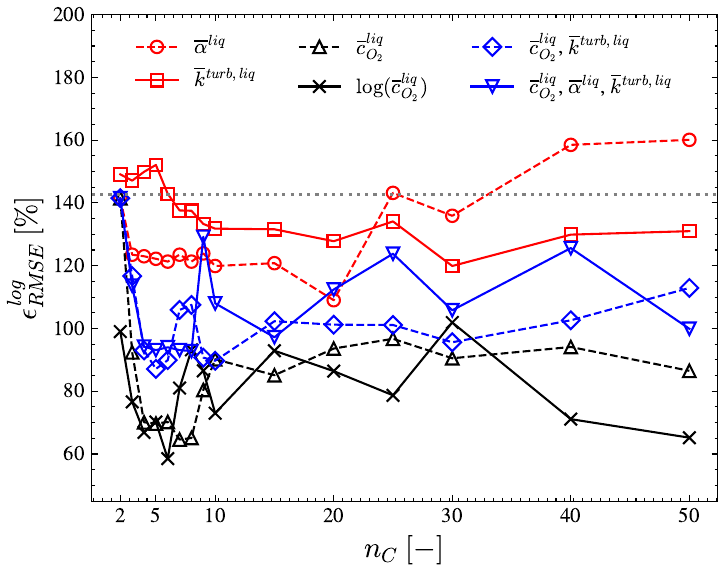}
            \caption{$\gls{error}_{RMSE}^{log}$}
            \label{fig:t1871_error_logRMSE}
        \end{subfigure}  
        \caption[Additional error results t1871]{Comparison of different error metrics for the  0.5th-Order Case. Total reaction rate error \eqref{eq:error_mol_rate} (a), \gls{rmse} \eqref{eq:error_rmse} (b), and logarithmic \gls{rmse} \eqref{eq:error_rmse_log} (c) for the 0.5\textsuperscript{th}-order case, using k-means clustering. The one-compartment conventional model yields a reaction rate error of $1.6\,[\%]$.}
        \label{fig:t1871_additional_results}
    \end{figure}

\autoref{fig:t1871_additional_results} shows the three error metrics. Notably, the one-compartment conventional model, as expected, overpredicts the reaction performance (by $1.6\,[\%]$). The compartment model slightly underpredicts the simulation results. Furthermore, the error increases with the number of clusters, a consequence of insufficient mixing: slightly elevated concentrations arise in regions of high mass transfer due to the absence of turbulent transport, reducing the effective driving concentration difference.

Neglecting turbulent transport causes concentration to accumulate artificially within smaller clusters. As cluster size increases, this effect is partially mitigated due to improved averaging and more adequate representation of mixing. Notably, the five-cluster case yields regions of accurate concentration prediction, predominantly at cluster centres, whereas the sixteen-cluster case exhibits systematic concentration discrepancies across large portions of the reactor.

    \begin{figure}[H]
        \centering
        \begin{subfigure}[h]{0.40\columnwidth}
            \centering
            \includegraphics[width=\linewidth]{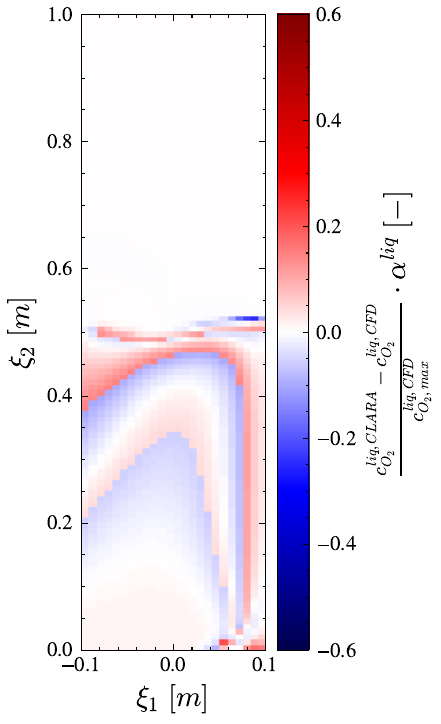}
            \caption{5 Clusters}
            \label{fig:t1871_error_field_c5}
        \end{subfigure}%
        \hfill %
        \begin{subfigure}[h]{0.40\columnwidth}
            \centering
            \includegraphics[width=\linewidth]{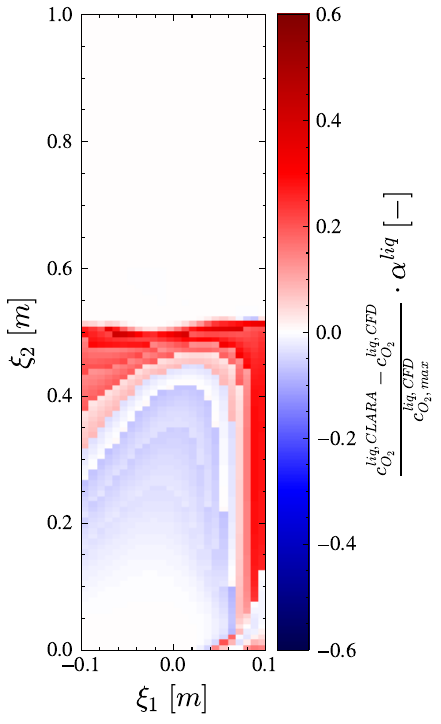}
            \caption{16 Clusters}
            \label{fig:t1871_error_field_c15}
        \end{subfigure}%

        \caption[Spatial error distribution for case t1871]{Spatial distribution of the local concentration error in each \gls{cfd} cell for 5 clusters (a) and 16 clusters (b), using k-means clustering with $\overline{\gls{conc}}^{\gls{liq}}_{O_2}$ as the clustering feature.}
        \label{fig:t1871_error_field_compairison_c5_c15}
    \end{figure}

\section{Software Repository}\label{app:supplimentary_material}

The \gls{clara} software repository can be cloned or downloaded from \href{https://gitlab.tugraz.at/simSci/clara-public.git}{the CLARA-Public gitlab repository}. Furthermore, the complete set of simulation cases and further supplementary material is accessible via the \href{https://cloud.tugraz.at/index.php/s/pqXN9LG6NggzxTN}{Supplementary Material}.

\end{document}

\endinput